\documentclass[journal,comsoc]{IEEEtran}

\usepackage[T1]{fontenc}
\usepackage{amsmath}
\usepackage{graphicx}
\normalsize
\IEEEoverridecommandlockouts
\usepackage{xcolor,cite,etoolbox}
\usepackage{algpseudocode}
\usepackage{color}
\usepackage{etoolbox}
\usepackage{cite}
\usepackage{hyperref}
\usepackage{booktabs,caption}
\usepackage[flushleft]{threeparttable} 
\usepackage[T1]{fontenc}
\usepackage[utf8]{inputenc}
\usepackage{textcase}
\usepackage{soul}
\usepackage{changes}
\usepackage{gensymb}
\usepackage{cite}
\usepackage{textcomp}
\usepackage{graphicx}
\usepackage{amssymb}
\usepackage{amsfonts}
\usepackage{epsfig}
\usepackage{color}
\usepackage{fancybox}
\usepackage{textcomp}
\usepackage{multirow}
\usepackage{setspa ce}
\usepackage{psfrag}
\usepackage{booktabs}
\usepackage{float}
\usepackage{mathrsfs}
\usepackage{amsmath}               
\usepackage[caption = false]{subfig}
\usepackage{caption}
\usepackage{algorithm}
\usepackage{varwidth}
\newtheorem{assumption}{Assumption}

\newtheorem{Remark}{Remark}

    \def\Complex{{\rm\rule[.23ex]{.03em}{1.1ex}\kern-.3em{C}}}

    \newcommand{\be}{\begin{equation}} \newcommand{\ee}{\end{equation}}
    \newcommand{\bea}{\begin{eqnarray}} \newcommand{\eea}{\end{eqnarray}}
    \newcommand{\benum}{\begin{enumerate}} \newcommand{\eenum}{\end{enumerate}}

        \newcommand{\qb}{{\bf b}}
        \newcommand{\qc}{{\bf c}}

        \newcommand{\qh}{{\bf h}}

        \newcommand{\qq}{{\bf q}}
        \newcommand{\qr}{{\bf r}}

        \newcommand{\qu}{{\bf u}}
        
        \newcommand{\qw}{{\bf w}}
        \newcommand{\qx}{{\bf x}}
        \newcommand{\qy}{{\bf y}}

        \newcommand{\qH}{{\bf H}}
        \newcommand{\qI}{{\bf I}}

        \newcommand{\qQ}{{\bf Q}}

        \newcommand{\qW}{{\bf W}}

        \newcommand{\qSigma}{{\boldsymbol \Sigma}}

        \newcommand{\qvarepsilon}{{\boldsymbol \varepsilon}}

        \newcommand{\qrho}{{\boldsymbol \rho}}

        \newcommand{\calN}{{\mathcal N}}

        \newcommand{\Ex}{{\sf E}}

        \newcommand*{\argmin}{\operatornamewithlimits{argmin}\limits}
        \newcommand*{\argmax}{\operatornamewithlimits{argmax}\limits}

\def\BibTeX{{\rm B\kern-.05em{\sc i\kern-.025em b}\kern-.08em
    T\kern-.1667em\lower.7ex\hbox{E}\kern-.125emX}}
\makeatletter 
\pretocmd\@bibitem{\color{black}\csname keycolor#1\endcsname}{}{\fail}
\newcommand\citecolor[1]{\@namedef{keycolor#1}{\color{blue}}}
\makeatother
\usepackage[cmintegrals]{newtxmath}
\begin{document}

\title{A Bayesian Receiver  with Improved Complexity-Reliability Trade-off in Massive MIMO Systems}

\author{Alva~Kosasih~\IEEEmembership{Student Member,~IEEE,} Vera~Miloslavskaya, Wibowo~Hardjawana~\IEEEmembership{Member,~IEEE,}  Changyang~She~\IEEEmembership{Member,~IEEE,} Chao-Kai~Wen~\IEEEmembership{Member,~IEEE,} and Branka~Vucetic,~\IEEEmembership{Life~Fellow,~IEEE}  
\thanks{This paper was published in part at the 2020 IEEE Wireless Communications and Networking Conference (WCNC) \cite{LBL_WCNC}. }
\thanks{ A. Kosasih, V. Miloslavskaya, W. Hardjawana, C. She, and B. Vucetic are with Centre of Excellence in Telecommunications, the University of Sydney,  Australia. (e-mail: $\{$alva.kosasih,vera.miloslavskaya$\}$@sydney.edu.au, $\{$wibowo.hardjawana,changyang.she,branka.vucetic$\}$@sydney.edu.au).}
\thanks{C. K. Wen is with the Institute of Communications Engineering, National Sun Yat-sen University, Kaohsiung, Taiwan (e-mail:  chaokai.wen@mail.nsysu.edu.tw).}}


\maketitle

\begin{abstract}
The stringent requirements on reliability and processing delay in the fifth-generation ($5$G) cellular networks  introduce  considerable challenges in the design of massive multiple-input-multiple-output (M-MIMO) receivers. The two main components of an M-MIMO receiver are a detector and a  decoder. To improve the trade-off between reliability and complexity, a Bayesian concept has been considered as a promising approach that enhances classical  detectors, e.g. minimum-mean-square-error detector. This work proposes an iterative M-MIMO detector based on a Bayesian framework, a  parallel interference cancellation scheme, and a decision statistics combining concept. We then develop a high performance M-MIMO receiver,  integrating the proposed detector with a  low complexity sequential decoding for polar codes.  Simulation results of the proposed detector show a significant performance gain compared to other low complexity detectors.  Furthermore, the proposed M-MIMO receiver with sequential decoding ensures one order magnitude lower complexity compared to  a receiver with stack successive cancellation decoding for polar codes from the 5G New Radio standard.
\end{abstract}

\begin{IEEEkeywords}
Massive MIMO,  5G receiver, Bayesian detector, PIC, DSC, Polar code, low complexity.
\end{IEEEkeywords}

\IEEEpeerreviewmaketitle

\section{Introduction}

Massive multiple-input-multiple-output (M-MIMO) technology plays a prominent role in the current wireless systems in increasing the number of  connections \cite{Overview.M.MIMO} and the spectral efficiency \cite{MIMO.5G} by using a large number of receive antennas. As the number of receive antennas increases, the  complexity of symbol detection at an M-MIMO receiver  increases as well, leading to a long processing delay. Essentially, there is a trade-off between the detection reliability and the processing delay. Improving the fundamental trade-off between reliability and processing delay is crucial for the fifth generation (5G) cellular networks \cite{3GPP_2017_TR,she2020deep}, where the requirement on the bit error rate (BER) lies in the range of $10^{-7}-10^{-5}$  and the processing delay should be lower than the duration of each  transmission time interval (TTI) ranging from $0.125$~ms to $1$~ms in 5G New Radio \cite{M-MIMO_URLLC_2018,3GPP2017Agree}.

\subsection{Related Works}

The state-of-the-art M-MIMO detectors can be classified into two categories: classical and Bayesian detectors. While optimal classical detectors, i.e. maximum likelihood (ML) detector \cite{ML},  can achieve a high detection reliability, such brute force algorithm suffers from a high complexity induced by searching for the ML combination of the transmitted symbols leading to a long processing time. To relax the high complexity, non-iterative \cite{LMMSE,J.Hoydis2013} and iterative \cite{MMSE_PIC,MMSE_SIC2,2019Albrem_WCNC_AppMatrxInv,2019Albreem_Survey_Detection,2020ZDan_ICICN_HybridMatrxInv,2020Chataut_DCOSS_HFADMM,PIC2001,Branka_PIC_book,PIC2012} classical detectors have been proposed.
 The minimum-mean-square-error (MMSE) detector \cite{LMMSE} is the best non-iterative classical detector in terms of the BER that has a moderate complexity. The reliability of the MMSE detector can be  improved by integrating the MMSE detector with iterative interference cancellation (IC) scheme. Specifically, MMSE detectors with parallel IC (PIC) and successive IC (SIC)  were proposed in \cite{MMSE_PIC} and \cite{MMSE_SIC2}, respectively. The MMSE-SIC provides a better BER performance compared to the MMSE-PIC detector at the expense of an increased processing delay. However, the computational complexity of the MMSE based detectors increases polynomially with the number of receive antennas due to the matrix inversion operation. As a result, the complexity of the MMSE based detectors is prohibitive for applications with a large number of receive antennas such as M-MIMO systems.

Considerable efforts have been made to develop detectors that can approach the MMSE performance with a much lower complexity\footnote{Due to the space limitation, we do not provide the details of approximate MMSE detectors except for the PIC-DSC detector scheme, which we use to cancel the user interference in the proposed detector. However, interested readers could see their details in \cite{2019Albrem_WCNC_AppMatrxInv,2019Albreem_Survey_Detection,2020ZDan_ICICN_HybridMatrxInv,2020Chataut_DCOSS_HFADMM}.}. In  \cite{2019Albrem_WCNC_AppMatrxInv} and \cite{2019Albreem_Survey_Detection}, the approximate matrix inverse (AMI) detectors based on  the conventional inverse approximations such as Neumann series, Richardson, successive-over relaxation (SOR), Gauss-Seidel (GS), optimized coordinate descent (OCD), Jacobi, and conjugate gradient (CG) methods  have been investigated. Latterly, to further reduce the complexity,  a combination of Jacobi and CG methods, referred to as the hybrid iterative (HI) detector, was proposed in  \cite{2020ZDan_ICICN_HybridMatrxInv}.
 A low complexity detector \cite{2020Chataut_DCOSS_HFADMM}, referred to as the Huber fitting based alternating direction method of multipliers (HF-ADMM) detector, was shown to provide near MMSE performance. Several low complexity detectors \cite{PIC2001,Branka_PIC_book,PIC2012}, referred to as the PIC-DSC detectors,  can achieve a near MMSE performance but offer a linear complexity with respect to the number of receive antennas. They first perform the matched filtering, i.e. multiplication of the received signals with  the conjugate transpose of the channel matrix, then estimate the interfering symbols using a decision statistics combining (DSC) scheme, and finally subtract these symbols from the received signals in a parallel manner to recover the symbols.  The latter is equivalent to the PIC scheme. Note that the performance of the MMSE detectors and their approximations is still far from the optimal ML detector.

More recently, the Bayesian M-MIMO detectors have been proposed to achieve a near ML performance by incorporating detection probability measures when estimating  symbols from the received signals  \cite{JGold_2011_GTA,Jespedes-TCOM14 ,EP-CG2017,2019GYao_Acc_LowCompEP,EP-NSA2018,EP_NSA_2020,EP2019,A.Kosasih,LAMA_Paper,LAMA_conf,Ma-17ACCESS}. The Bayesian framework is used to satisfy the maximum a posteriori (MAP) criterion for M-MIMO detection by using the Bayesian rule, i.e. $p(\qx|\qy) = {p(\qy|\qx) p(\qx)}/{p(\qy)}$ \cite{JGold_2011_GTA}. The best Bayesian detector in terms of the BER performance is the expectation propagation (EP) detector \cite{Jespedes-TCOM14 }. It significantly outperforms the MMSE-SIC detector and can achieve a near ML performance. The EP detector uses the Bayesian framework with the MMSE scheme as its interference suppressor. The EP detector requires  to perform a matrix inversion operation in every iteration which results in a polynomial growth of the complexity with the number of receive antennas. 

 Low-complexity versions of the EP detector have been proposed in \cite{EP-CG2017,2019GYao_Acc_LowCompEP,EP-NSA2018,EP2019,A.Kosasih,EP_NSA_2020}. The matrix inversion operations of the original EP detector have been approximated by the CG, Sherman-Morrison formula, and  Neumann series approximation (NSA)  in EP-CG  \cite{EP-CG2017}, EP-SU \cite{2019GYao_Acc_LowCompEP},  and  EP-NSA \cite{EP-NSA2018,EP_NSA_2020}  detectors, respectively.  Nevertheless, the EP-CG and EP-SU have a high processing delay due to requiring a large number of iterations \cite{2019Albrem_WCNC_AppMatrxInv} and the nature of the sequential updating, respectively. In \cite{EP2019}, the EP approximation (EPA) detector was proposed,  where the complexity reduction is achieved by exploiting channel hardening phenomenon. 
Another recent low complexity EP detector \cite{A.Kosasih}, referred to as the decentralized EP (D-EP) detector, divides receive antennas into clusters, where each cluster independently processes the local received signal, thereby naturally reducing the size of matrix inversion. 
However, none of \cite{EP-CG2017,2019GYao_Acc_LowCompEP,EP-NSA2018,EP_NSA_2020,EP2019,A.Kosasih} can achieve a near EP performance without performing any matrix inverse operation or its approximation, and thus these detectors are unable to achieve a linear complexity with respect to the number of receive antennas.
The only existing low complexity Bayesian based detector that has such linear complexity is the approximate message passing (AMP) \cite{LAMA_Paper,LAMA_conf}  detector.  However, the BER achieved by the AMP detector is much worse than the  EP detector. The enhanced AMP detector \cite{Ma-17ACCESS},  referred to as the  orthogonal AMP (OAMP) detector, outperforms the AMP detector in medium scale MIMO systems. Nevertheless, it needs to compute matrix inversion operations, resulting in a high complexity. To the best of authors' knowledge, there is no Bayesian based detector able to  achieve a near ML BER with a linear complexity with respect to the number of receive antennas.

To develop a high performance M-MIMO receiver, channel coding has been incorporated into the M-MIMO system. We focus on polar codes, which
have been recently adopted by the New Radio standard \cite{polarNR2018} for the control channels in the 5G enhanced mobile broadband (eMBB) scenario. Polar codes are likely candidates for future 5G and beyond standards. In particular, they are suitable for small packet transmission in ultra-reliable and low-latency communications (URLLC) \cite{Sharma2019PolarCA}. Polar codes are able to satisfy the demands of 5G systems such as maintaining good error-correction performance over diverse code and channel parameters \cite{Bioglio2020DesignOP}. Recursive structure of polar codes enables decoding using the successive cancellation (SC) algorithm \cite{arikan2009channel}. A major drawback of the SC algorithm is that it cannot correct errors which may occur at early phases of the decoding process. This problem is solved in the SC list (SCL) \cite{tal2011list}  by keeping  a  list  of  the  most  probable  paths within the  code  tree. The SC stack (SCS) \cite{niu2012crcaided}  has the same error rate performance as the SCL, but a much lower complexity due to avoiding the search over many useless paths. If the information about distribution of the input noisy symbols is provided, the complexity could be further reduced by the sequential decoding algorithm  \cite{miloslavskaya2014sequential,Trifonov2018ASF}. State-of-the-art polar-coded MIMO systems \cite{Watanabe2017PerformanceOP} and \cite{Dai2018PolarCodedMS} are based on the SC, SCL and SCS decoding. To the best of authors' knowledge, no results on the polar-coded M-MIMO receiver with the sequential decoder or the Bayesian detector have been reported.

\subsection{Contributions}

We propose a novel Bayesian M-MIMO detector,  referred to as the Bayesian PIC-DSC (B-PIC-DSC) detector,  to improve the trade-off between reliability and complexity  in 5G networks. The developed detector consists of three modules: Bayesian symbol observation (BSO), Bayesian symbol estimator (BSE), and DSC modules. They are used to calculate the probability density functions (pdfs) of the symbols by using the matched filter based PIC scheme, calculate the Bayesian symbol estimates, and refine the symbols by utilizing the DSC concept, respectively. We further develop a high performance low complexity polar coded M-MIMO receiver by integrating the proposed detector with the sequential decoder for polar codes.
The main contributions of this paper are summarized as follows.
\begin{itemize}
\item We develop a new Bayesian M-MIMO detector that can achieve a near ML performance. The complexity increases linearly with the number of receive antennas by employing the matched filter based PIC-DSC  scheme \cite{Branka_PIC_book} and the Bayesian concept \cite{Jespedes-TCOM14}. The performance of the proposed detector is evaluated not only when the M-MIMO channels are independent and identically distributed (i.i.d.) and perfectly known, but also in the presence of channel estimation error and spatial correlation between receive antennas, which could be the main performance degradation factors in practical M-MIMO systems. The simulation results show that with the ratio of transmit-to-receive antennas of up to $50 \%$, the B-PIC-DSC detector can achieve a near ML performance.
\item We derive a closed-form BER approximation of the proposed M-MIMO detector. The closed-form expression can be used to predict the BER performance of the proposed detector without extensive simulations. Our simulation results show that the approximation is accurate for various ratios of transmit-to-receive antennas.
\item We integrate the proposed detector with a low complexity sequential decoder for polar codes by using a closed-form variance approximation of the output log-likelihood ratios. The approximation is derived by leveraging the evolution technique in \cite{XYuan2008}. The sequential decoder employs the variance approximation to reduce the search space. Simulation results demonstrate that the proposed detector with the sequential decoding ensures a significant reduction in the complexity compared to detectors integrated with the SCS/SCL decoder in case of the polar codes from the 5G New Radio standard. 
\end{itemize}

This paper is organized as follows. The system model for uncoded M-MIMO systems is introduced in Section II. It is followed by the details of the B-PIC-DSC detector, in Section III. The closed-form BER approximation  and the complexity analysis of the B-PIC-DSC detector are presented in Section IV and Section V, respectively. Section \ref{sPolar} describes a low complexity polar-coded M-MIMO receiver based on the B-PIC-DSC detector.  In Section VII, we provide in-depth performance evaluations. Finally, Section VIII concludes the paper. 

\subsection{Notations}

We first define the notations used in the paper.  $\mathbf{I}$ denotes an identity matrix.  For any matrix $\mathbf{A}$, the notations $\mathbf{A}^{T}$, $\mathbf{A}^{H}$ and ${\sf tr}(\mathbf{A})$ stand for transpose, conjugate transpose and trace of $\mathbf{A}$, respectively. 
  $\|\qq\|$ denotes the Frobenius norm of vector $\qq$.   $q^*$ denotes the complex conjugate of a complex number $q$. 
 ${\Ex}[\qx]$ is the mean of random vector $\qx$ and ${\mathrm{Var} }[\qx] = {\Ex}\big[\left(\qx-{\Ex}[\qx]\right)^2\big]$ is its variance.   $\calN(x_k,c_k;v_k)$ represents a complex single variate Gaussian distribution  for a  random variable $x_k$ with mean $c_k$ and variance $v_k$. Let $\qx = [x_1, \cdots, x_K]^T$ and $\qc = [c_1, \cdots, c_K]^T$.  A multivariate Gaussian distribution for a random vector $\qx$ is denoted by $\calN(\qx,\qc;\qSigma^{(t)}) $, where 	$\qc$ is a mean vector and $\qSigma^{(t)}$ is a covariance matrix.

\begin{figure}
\centering
{\includegraphics[scale=0.33]{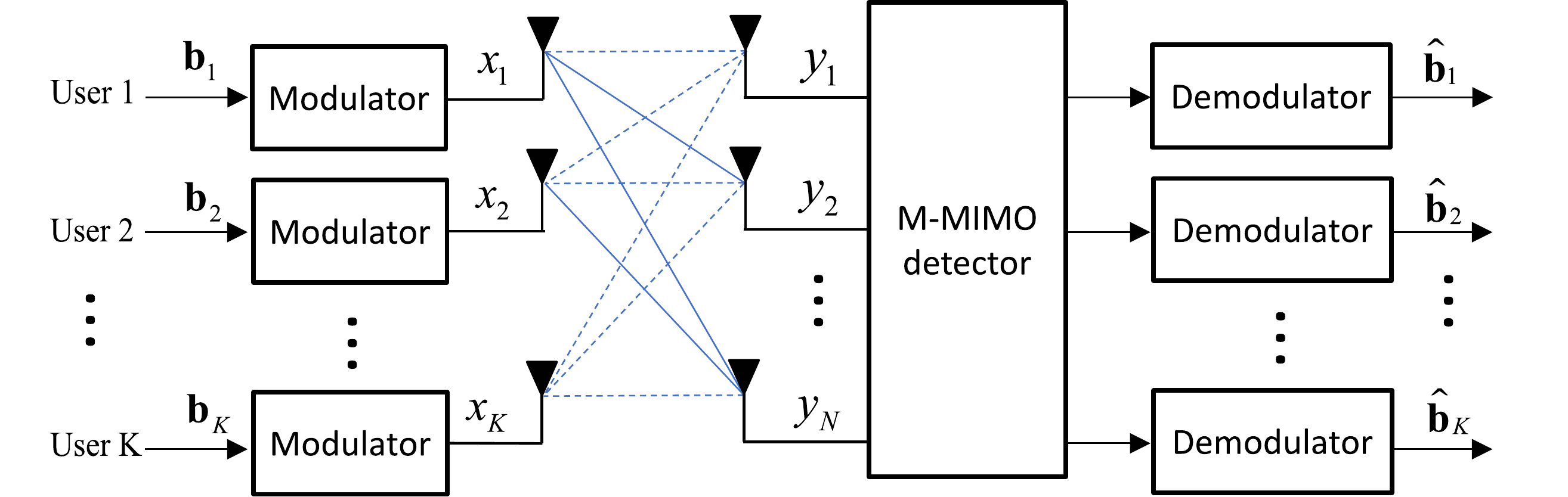}}
\caption{The uplink M-MIMO system}
\label{up_M-MIMO}
\end{figure}

\section{System Model}

We consider an  uncoded M-MIMO system used to transmit information streams generated by $K$ single-antenna users, as depicted in  Fig. \ref{up_M-MIMO}. The M-MIMO system includes a modulator located on the user side and an M-MIMO detector together with a demodulator located on the base station. The base station is equipped with  a large number of receive antennas $N>>K$ to simultaneously serve the users. 
User $k$ maps $\log_2(M)$ bits of its information stream $\qb_k$ to a symbol $x_k \in \Omega$  using a quadrature amplitude modulation (QAM) technique, where $\Omega = [ s_1, \dots, s_M]$ is a constellation set of M-QAM  and $s_m$ is one of the constellation points. The transmitted symbols are uniformly distributed and the received signal is given by 
\begin{equation} \label{eII_1}
\qy = \qH \qx + \qvarepsilon,
\end{equation}
where $\qx = [x_1, \cdots, x_K]^T$, $\qy=[y_1, \ldots, y_N]^{T}$,  $\qH=[\qh_1, \ldots, \qh_K]  \in \mathbb{C}^{N\times K} $ is the coefficient  matrix of complex memoryless Rayleigh fading channels between $K$ transmit antennas and $N$ receive antennas,  $\qh_k$ is the $k$-th column vector of matrix $\qH$ that denotes wireless channel coefficients between receiver antennas and user $k$, where each coefficient follows a Gaussian distribution with zero mean and unity variance, and $\qvarepsilon\in \mathbb{C}^N$ denotes the additive white Gaussian noise (AWGN) with a zero mean and covariance matrix $\sigma^2 \qI$. The SNR of the system is defined as SNR $ = 10 {\sf{log}}_{10} \left( \frac{K \mathit{E}_s}{\sigma^2} \right) $ dB, where $\mathit{E}_s$ is the energy per transmit antenna. We normalize the total transmit energy such that $K \mathit{E}_s =1 $. Given a received vector $\qy\in\mathbb{C}^N$, the optimal detector using the MAP decision rule, to find 
\begin{flalign}\label{MAP}
\hat{\qx} &= \argmax_{\qx \in \Omega^K} p(\qx|\qy) \notag \\
& = \argmin_{\qx \in \Omega^K} \| \qy-\qH \qx\|^2.
\end{flalign}
Note that here the MAP detection rule reduces to the ML detection rule, since the transmitted symbols are uniformly distributed. The  complexity of the optimal detector in \eqref{MAP} grows exponentially  with the number of transmit antennas and the size of the constellation set \cite{ML}. Thus, it cannot be implemented in the practical M-MIMO systems. The MMSE detector is used to relax the complexity of the optimal detectors,  where the transmitted symbols are approximated as \cite{LMMSE}
\begin{equation}\label{LMMSE}
\hat{\qx} \approx \left( \qH^H\qH +\sigma^{2} \mathbf{I} \right)^{-1}\qH^H \qy.
\end{equation}
The matrix inversion operation used in \eqref{LMMSE} is still costly as its complexity increases polynomially with the number of receive antennas. 
In contrast to the MMSE scheme, the iterative matched filter based PIC scheme avoids the matrix inversion operations. Specifically, the estimation of the symbol of user $k$ in iteration $t$, $x_{{\rm PIC},k}^{(t)}$, is given in \cite{PIC2001,Branka_PIC_book} as 
\begin{equation}\label{eA1_a02}
x_{{\rm PIC},k}^{(t)} = \frac{\qh_k^H \left( \qy-\qH \qx_{{\rm PIC} \backslash k}^{(t-1)} \right)}{\| \qh_k \|^2},
\end{equation} 
where $ \qx_{{\rm PIC} \backslash k}^{(t-1)}   = \left[x_{{\rm PIC}, 1}^{(t-1)}  , \dots, x_{{\rm PIC}, k-1}^{(t-1)} , 0, x_{{\rm PIC}, k+1}^{(t-1)}, \dots, x_{{\rm PIC}, K}^{(t-1)} \right]^T $ are the estimated symbols in the $(t-1)$-th iteration.  At the first iteration, we initialise $\qx_{{\rm PIC}, k}^{(0)}  =\bold{0}$ indicating that the PIC is inactive and thus we have $  x_{{\rm PIC},k}^{(1)} =\frac{\qh_k^H\qy }{\|\qh_k\|^2} $, which is the expression of the matched filter  \cite{MRC-LDPC}.

\section{Bayesian PIC M-MIMO Detector}
\label{sDetector}

\begin{figure*} 
\centering
\includegraphics[scale=0.43]{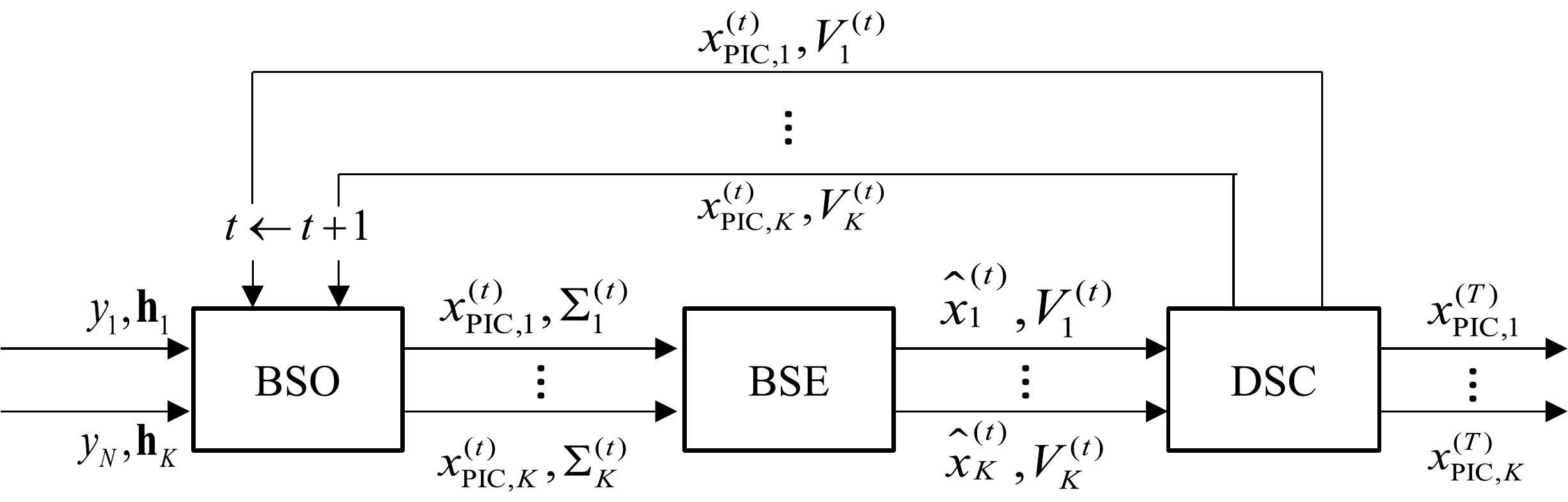}
\caption{The architecture of the B-PIC-DSC detector}\label{F0}
\end{figure*}

In this section, we propose a novel Bayesian PIC-DSC detector, referred to as the B-PIC-DSC detector, to be employed in an uplink M-MIMO system,  illustrated in  Fig. \ref{up_M-MIMO}. The structure of the B-PIC-DSC detector is shown in Fig. \ref{F0}. It consists of three modules: a BSO module that computes the pdfs of the transmitted symbols from the  received signals by using  the matched filter based PIC scheme; a BSE module that obtains the Bayesian symbol estimates based on the computed pdfs; and a DSC module that refines the transmitted symbol estimates by using the outputs of the BSE module and returns the refined symbols to the BSO module.

\subsection{Bayesian Symbol Observation}

The computation of the pdfs of symbols from the received signals is done by treating  $\qx$ in \eqref{eII_1} as a random vector. According to Bayesian rule, the posterior probability of the transmitted symbols $\qx$, given the received signals $\qy$, can be expressed as follows 
\begin{equation}\label{Bayesian_rule}
p(\qx|\qy) = \frac{p(\qy|\qx) p(\qx)}{p(\qy)},
\end{equation}
where $p(\qy|\qx) =\mathcal{N}\left( \qy, \qH\qx; \sigma^2\qI \right)$. Since the transmitted symbols are uniformly distributed, $p(\qx|\qy)$  in \eqref{Bayesian_rule} can be simplified as 
\begin{equation}\label{dist_of_x_y}
p(\qx|\qy) \propto \mathcal{N}\left( \qy, \qH\qx; \sigma^2\qI \right).  
\end{equation}
Obtaining symbol estimates by using the MAP criterion  \eqref{MAP} with $p(\qx|\qy)$ from  \eqref{dist_of_x_y} is an NP hard problem. To address this, we employ the Bayesian posterior approximation, which allows us to iteratively approximate the posterior function $p(\qx|\qy)$ by a product of independent Gaussian functions\cite{JGold_2011_GTA}, denoted as $\hat{p}^{(t)}(x_k|\qy)$ in iteration $t$. 
Accordingly, we approximate $p(\qx|\qy) $  as
\begin{equation}\label{iid_assumption}
p(\qx|\qy)   \approx \prod_{k=1}^K \underbrace{\mathcal{N}\left( x_k, x_{{\rm PIC},k}^{(t)}; \Sigma_k^{(t)} \right)}_{\hat{p}^{(t)}(x_k|\qy)},
\end{equation}
where $x_{{\rm PIC},k}^{(t)}$  is the soft estimate  of $x_k$ in iteration $t$, which is given in \eqref{eA1_a02}, since we use the matched filter based PIC scheme to detect the  transmitted  symbols. The variance $ \Sigma_k^{(t)} $ of the  $k$-th PIC symbol estimate  is derived as
\begin{align}\label{eA1_a01}
\Sigma_k^{(t)} & =  \frac{1}{\left( \sum_{n=1}^{N}h_{n,k}^* h_{n,k}\right)^2} \left( \sum_{\substack{j=1 \\  j\neq k}}^{K} s_j^2 V_j^{(t-1)} +  \sum_{n=1}^N  \left( h_{n,k}^*h_{n,k} \right) \sigma^2 \right) \notag \\
& \approx \frac{\sigma^2}{\sum_{n=1}^{N}h_{n,k}^* h_{n,k}},
\end{align}
where the approximation is taken from \cite{LBL_WCNC},  $s_j =\sum_{n=1}^{N} h_{n,k}^*h_{n,j}$, and $V_j^{(t-1)}$ is the variance of the Bayesian symbol estimate in  iteration $t-1$.  The derivation of \eqref{eA1_a01} can be found in Appendix \ref{Ap.Var_PIC}.  The posterior approximations,  $\hat{p}^{(t)}(x_k|\qy)={\cal{N}}\left( x_k,x^{(t)}_{{\rm PIC},k}; {\Sigma}^{(t)}_k \right), k=1, \dots, K,$  are then forwarded to the BSE module, as shown in Fig. \ref{F0}.
  
\subsection{Bayesian Symbol Estimator}
      
In the BSE module, we compute the Bayesian symbol estimate of the $k$-th user by using $\hat{p}^{(t)}(x_k|\qy)$ obtained from the BSO module.  
Since $p(\qx|\qy)$ can be factorized according to \eqref{iid_assumption}, criterion in  \eqref{MAP} can be decomposed as 
\begin{flalign}\label{MAP_component}
\hat{x}_k^{(t)} &= \arg \max_{a \in \Omega} \hat{p}^{(t)}(x_k=a|\qy).
\end{flalign}
Note that with the Bayesian framework, we can approximate the exponentially complex MAP criterion in \eqref{MAP} with the expression in \eqref{MAP_component}, which has a linear complexity. 
The Bayesian symbol estimate and its variance  are respectively given in \cite{Probability} as
\begin{equation}\label{eA1_b01}
\hat{x}_k^{(t)} =\Ex\left[x_k \Big| x_{{\rm PIC}, k}^{(t)} ,\Sigma^{(t)}_k \right] =\sum_{a \in \Omega} a  \hat{p}^{(t)}{\left(x_k=a|\qy\right)}
\end{equation}
\begin{equation}\label{eA1_b02}
V_k^{(t)}=\mathrm{Var} \left[x_k \Big| x_{{\rm PIC}, k}^{(t)} ,\Sigma^{(t)}_k  \right]  =\Ex  \left[ \left| x_k  - \Ex\left[x_k \Big| x_{{\rm PIC}, k}^{(t)} ,\Sigma^{(t)}_k \right] \right|^{2} \right], 
\end{equation}
where $\hat{p}^{(t)}\left(x_k|\qy\right)$ is normalized so that  $\sum_{a\in \Omega}\hat{p}^{(t)}\left(x_k=a|\qy\right) =1$.
The outputs of the BSE module, $\hat{x}_k^{(t)} $ and $V_k^{(t)} $, $k= 1, \dots, K$, are then sent to the DSC module.

\subsection{Decision Statistics Combining}
In the matched filter based PIC scheme, the interference canceller is inactive in the first iteration and thus the inter-symbol interference is very high. From the second iteration, the symbol estimates approach the corresponding transmitted symbols as the interference is gradually mitigated. Consequently, the value of $\hat{x}_k^{(t)}$ varies significantly in the first few iterations  and hence the correlation between $\hat{x}_k^{(t)}$ and $\hat{x}_k^{(t-1)}$ is low when $t$ is small, as will be shown in Section \ref{Sec:app2}. Such a feature can be exploited to increase the diversity of symbol estimates by forming decision statistics,  referred to as the DSC  \cite{PIC2001,Branka_PIC_book}. The decision statistics consist of a linear combination of the symbol estimates in two consecutive iterations according to \cite{Branka_PIC_book}
\begin{equation}\label{DSC}
x_{{\rm DSC},k}^{(t)} = \left( 1-\rho_{{\rm DSC},k}^{(t)} \right)  \hat{x}_k^{(t-1)}  +   \rho_{{\rm DSC},k}^{(t)}   \hat{x}_k^{(t)}
\end{equation}
\begin{equation}\label{DSC_Var}
V_{{\rm DSC},k}^{(t)} = \left( 1-\rho_{{\rm DSC},k}^{(t)} \right)  V_k^{(t-1)}  +   \rho_{{\rm DSC},k}^{(t)}   V_k^{(t)}.
\end{equation}
As illustrated in Fig. \ref{F0}, $x_{{\rm DSC},k}^{(t)}$ and $V_{{\rm DSC},k}^{(t)} $ are computed in the DSC module. 

Existing Bayesian detectors obtain the weighting coefficients for \eqref{DSC} and \eqref{DSC_Var}, from trial and error processes \cite{Jespedes-TCOM14,2019GYao_Acc_LowCompEP,LAMA_Paper,LAMA_conf,A.Kosasih}. The weighting coefficients vary over the system configurations. In this work, we leverage the DSC concept to avoid the trial and error processes. Specifically, the weighting coefficients in the linear combinations are determined by maximizing the signal-to-interference-plus-noise-ratio  (SINR)-variance. In iteration $t$, the $k$-th coefficient is given as \cite{Branka_PIC_book}
\begin{equation}\label{DSC_coef}
\rho_{{\rm DSC},k}^{(t)} =  \frac{e_k^{(t-1)}}{e_k^{(t)}+e_k^{(t-1)}}, 
\end{equation}
where $e_k^{(t)}$ is defined as the instantaneous square error of the $k$-th symbol estimate, which can be computed by using a linear filter such as matched or zero forcing (ZF) filter. That is,
\begin{flalign}\label{DSC_error}
 e_k^{(t)}  =  \left\| \qw_k^H \left(  \qy - \qH \hat{\qx}^{(t)} \right)\right\|^2,  
\end{flalign}
where $\qw_k$ is the $k$-th column vector of the linear filter for user $k$. For the B-PIC-DSC detector, we use the matched filter, therefore $\qw_k^H =\frac{\qh_k^H}{\| \qh_k\|^2}$. Nevertheless, the performance of our detector is still close to the optimal detector.
The iterative process will stop if the following condition is satisfied,
\begin{equation}\label{eq_convergence}
 \|x_{{\rm DSC},k}^{(t)} - x_{{\rm DSC},k}^{(t-1)} \| \leq \zeta  \text{  or  } t = T_{\rm max}, 
\end{equation}
where  $\zeta$ is the minimum acceptable difference of $x^{(t)}_{{\rm DSC},k}$ in two consecutive iterations, and $T_{\rm max}$ is the maximum number of iterations. We then use $x_{{\rm DSC},k}^{(t)} $ and $V_k^{(t)} $  as the input of the BSO module by assigning the value of $x_{{\rm DSC},k}^{(t)}$ to $x_{{\rm PIC},k}^{(t)}$ and $ V_{{\rm DSC},k}^{(t)}$ to $V_k^{(t)} $,
\begin{subequations}\label{assign}
\begin{equation}
x_{{\rm PIC},k}^{(t)} \leftarrow x_{{\rm DSC},k}^{(t)}
\end{equation}    
\begin{equation}
V_k^{(t)} \leftarrow V_{{\rm DSC},k}^{(t)}.
\end{equation}
\end{subequations}
The complete pseudo-code is shown in Algorithm \ref{A1}.
\begin{algorithm}
\caption{B-PIC-DSC detector}
\label{A1}
\begin{algorithmic}[1]
\State {\textbf{Input: } $K, \Omega, \qy, \qH, \sigma^2, \hat{x}_{{\rm PIC},k}^{(0)} \leftarrow 0,  V_j^{(0)} \leftarrow 1 , T_{\rm max} \leftarrow 10 $}
\State {\textbf{Output: }  $\hat{\qx}^{(T)}$} 
	\For {$t=1,\dots, T_{\rm max}$}
		\For {$k=1,\dots, K$ (parallel execution)}
	    		\Statex \textbf{\quad\ \quad\, The BSO Module:}
			\State \begin{varwidth}[t]{0.84\linewidth}Compute the expected value of $x_k$ based on the PIC scheme, $x_{{\rm PIC},k}^{(t)}$  in \eqref{eA1_a02}\end{varwidth} 
			\State Compute the variance ${\Sigma}_k^{(t)}$  in \eqref{eA1_a01}
			\Statex \textbf{\quad\ \quad\, The BSE Module:}
			\State Compute  the Bayesian symbol estimate $\hat{x}_k^{(t)}$  in \eqref{eA1_b01}
			\State  Compute the Bayesian variance   $V_k^{(t)} $ in \eqref{eA1_b02}
			\Statex \textbf{\quad\ \quad\, The DSC Module:}
			\State \begin{varwidth}[t]{0.84\linewidth}Compute the instantaneous error approximate of the $k$-th symbol estimate $e_{k}^{(t)} $  in \eqref{DSC_error}\end{varwidth}
			\State Compute the DSC coefficient  $\rho_{{\rm DSC},k}^{(t)} $  in \eqref{DSC_coef} 
			\State \begin{varwidth}[t]{0.84\linewidth}Compute the weighted Bayesian symbol estimate $x_{{\rm DSC},k}^{(t)} $  in \eqref{DSC}\end{varwidth} 
			\State \begin{varwidth}[t]{0.84\linewidth}Compute the weighted Bayesian variance $V_{{\rm DSC},k}^{(t)}$ in \eqref{DSC_Var}\end{varwidth}  
			\State Compute \eqref{assign}  
		\EndFor
		\If {$ \|x_{{\rm DSC},k}^{(t)} - x_{{\rm DSC},k}^{(t-1)} \| \leq 10^{-4} $}
			\State {break}
		\EndIf
	\EndFor
		\State $T\leftarrow t$ 
\end{algorithmic}
\end{algorithm}

\subsection{The Improved B-PIC-DSC Detector}
\label{sImprDetector}

In the first iteration, the proposed B-PIC-DSC detector relies on the matched filter to estimate the symbols. When there is a large number of interfering symbols, the transmitted symbol estimates in the first iteration suffer from considerable detection errors, propagated along with the iterations, and eventually result in performance degradation. To improve the performance of the B-PIC-DSC detector, we propose an  improved B-PIC-DSC (IB-PIC-DSC) detector that applies the MMSE scheme in \eqref{LMMSE} only in the first iteration,
\begin{flalign}\label{improve_first_iter}
\qx_{\rm PIC}^{(0)}  &= (\qH^H\qH+\sigma^2\qI)^{-1} { \qH^H\mathbf{y}} \notag \\
&= \qW^H\qy.
\end{flalign}
The $k$-th row of the MMSE matrix $\qW^H$ denoted by $\qw^H_k$  is then used to calculate the approximation of instantaneous errors in \eqref{DSC_error}. For $t \geq 1$, the IB-PIC-DSC detector performs identical computations as the B-PIC-DSC detector. It is worth noting that the IB-PIC-DSC detector performs the inverse matrix operation only in the first iteration. This is different from the  EP   and MMSE-SIC detectors, which calculate the inverse  matrix operation in every iteration.

\section{BER Approximation of The Proposed Detectors}

In this section, we derive a closed-form BER approximation with respect to the ratio of transmit-to-receive antennas for the proposed detectors. With the BER approximation, one can calculate the number of users that  can be served simultaneously in order to achieve a certain BER requirement, provided a sufficient number of receive antennas on the receiver side. The approximation is derived by leveraging the SINR-variance evolution technique, proposed for systems based on the turbo principle by \cite{XYuan2014PA}. To simplify the analysis,  we first set the DSC coefficients in \eqref{DSC_coef} equal to unity. This implies that we do not exploit the diversity of symbol estimates when deriving our BER approximation. The recursive processes in the B-PIC-DSC detector can then be iteratively characterized by the SINR of the BSO module and the variance of the BSE module. 

\subsection{SINR in the BSO module}

When the numbers of receive antennas and users grow very large while the ratio of transmit-to-receive antennas  is fixed, the variance $\Sigma^{(t)}_k$ of the pdf $\hat{p}^{(t)}(x_k|\qy)$ have a similar value to   \cite{A.Kosasih} $v^{(t)}\triangleq \frac{1}{K}{\sf tr} \left( \qSigma^{(t)}\right),  k=1,\dots,K.$ 
The following assumption, originally suggested in  \cite{XYuan2014PA}, is employed to obtain the BER approximation for the proposed detector. 
\begin{assumption}
\label{assump:x}
The symbol estimates in the BSO module can be expressed as \cite{XYuan2014PA}
\begin{equation}\label{Awgn}
 x_{{\rm PIC},k}^{(t)} = x_k+\tilde{\varepsilon}^{(t)},
\end{equation}  
\end{assumption}
where $\tilde{\varepsilon}^{(t)}$ is a Gaussian random variable with mean $0$ and variance $v^{(t)}$. The SINR of the BSO module can be computed as $1/v^{(t)}$. In the last iteration, the SINR of the BSO module is used to calculate the final BER approximation.

In the first iteration, where the PIC scheme is inactive, the symbol estimates in the BSO module are computed based on the matched filter and MMSE schemes for the B-PIC-DSC and IB-PIC-DSC detectors, respectively. Accordingly, the  variances for both schemes  are given in  \cite{Ma-17ACCESS,J.Hoydis2013} as
\begin{flalign}\label{MSE-IB-PIC-DSC}
 v^{(0)}= \begin{cases}
\frac{\alpha \sigma^{2}+(\alpha-1)+   \sqrt{(\alpha \sigma^2+(\alpha-1))^2+4\alpha \sigma^2}}{2}, & \text{for the IB-PIC-DSC} \\
\frac{K-1+K\sigma^2}{N}, & \text{for the B-PIC-DSC},
\end{cases}
\end{flalign}
where $\alpha \triangleq  \frac{K}{N}$ is  the ratio of transmit-to-receive antennas.
From the second iteration onward, the variances of the B-PIC-DSC and IB-PIC-DSC detectors in the BSO module are identical and given as,
\begin{flalign}\label{SINR_BSO}
v^{(t)} &= \underbrace{\frac{K-1}{N}V_j^{(t-1)}}_{\text{multiple user interference}}+\underbrace{\frac{\sigma^2}{N}}_{\text{noise}}, t>0, 
\end{flalign}
where the derivation of \eqref{SINR_BSO} is given in Appendix B and $V_j^{(t-1)} $ is the mean square error (MSE) of the user's symbol defined as 
\begin{equation}\label{MSE_user}
V_j^{(t-1)} = \Ex [|x_j-\hat{x}_j^{(t-1)}|^2].
\end{equation}
 Note that when the numbers of transmit and receive antennas grow very large, all values of $V_j^{(t-1)}, j = 1,\dots,K $, will be almost identical. For notational simplicity, we replace the subscript $j$ in $V_j^{(t-1)}$ with $k$  in the following discussion, where $ k = 1,\dots,K $.

\begin{algorithm}
\caption{BER approximation of the proposed detectors \label{A3}}
\begin{algorithmic}[1]
\State \textbf{Input: } $  K, N,  \sigma^2$, $ V_j^{(0)} \leftarrow 1$
\State \textbf{Output: }  ${\rm  BER}_{\rm ap} $
\State  Compute $v^{(0)}$ in \eqref{MSE-IB-PIC-DSC} 
	\For {$t=1, 2, \dots, \infty $}
			\State Calculate $v^{(t)}$ in \eqref{SINR_BSO} 
			\State Calculate $V_k^{(t)}$ in \eqref{BSE_Analytical}
		\If {$ \|v^{(t)}-v^{(t-1)}\| \leq 10^{-4} $}
			\State {break}
		\EndIf
	\EndFor
\State {$T\leftarrow  t $} 
\State Compute the ${\rm  BER}_{\rm ap} $ in \eqref{BER_SE}

\end{algorithmic}
\end{algorithm}

\subsection{The variance of the output of the BSE module}

\begin{figure}
\centering
\subfloat[$N=128$ and $M=4$]
{\includegraphics[scale=0.32]{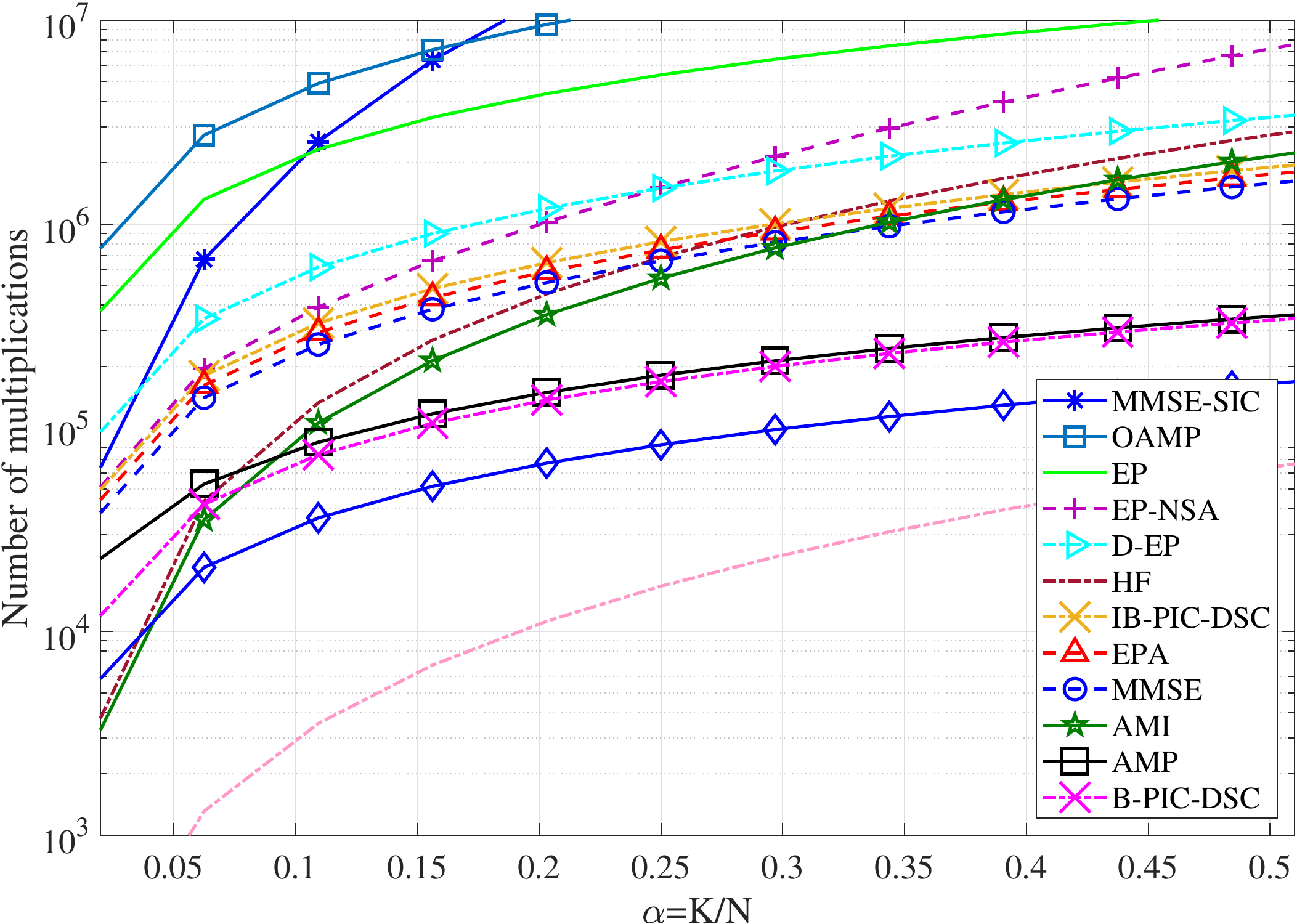}}\hfill
\centering
\subfloat[$N=256$ and $M=4$]
{\includegraphics[scale=0.32]{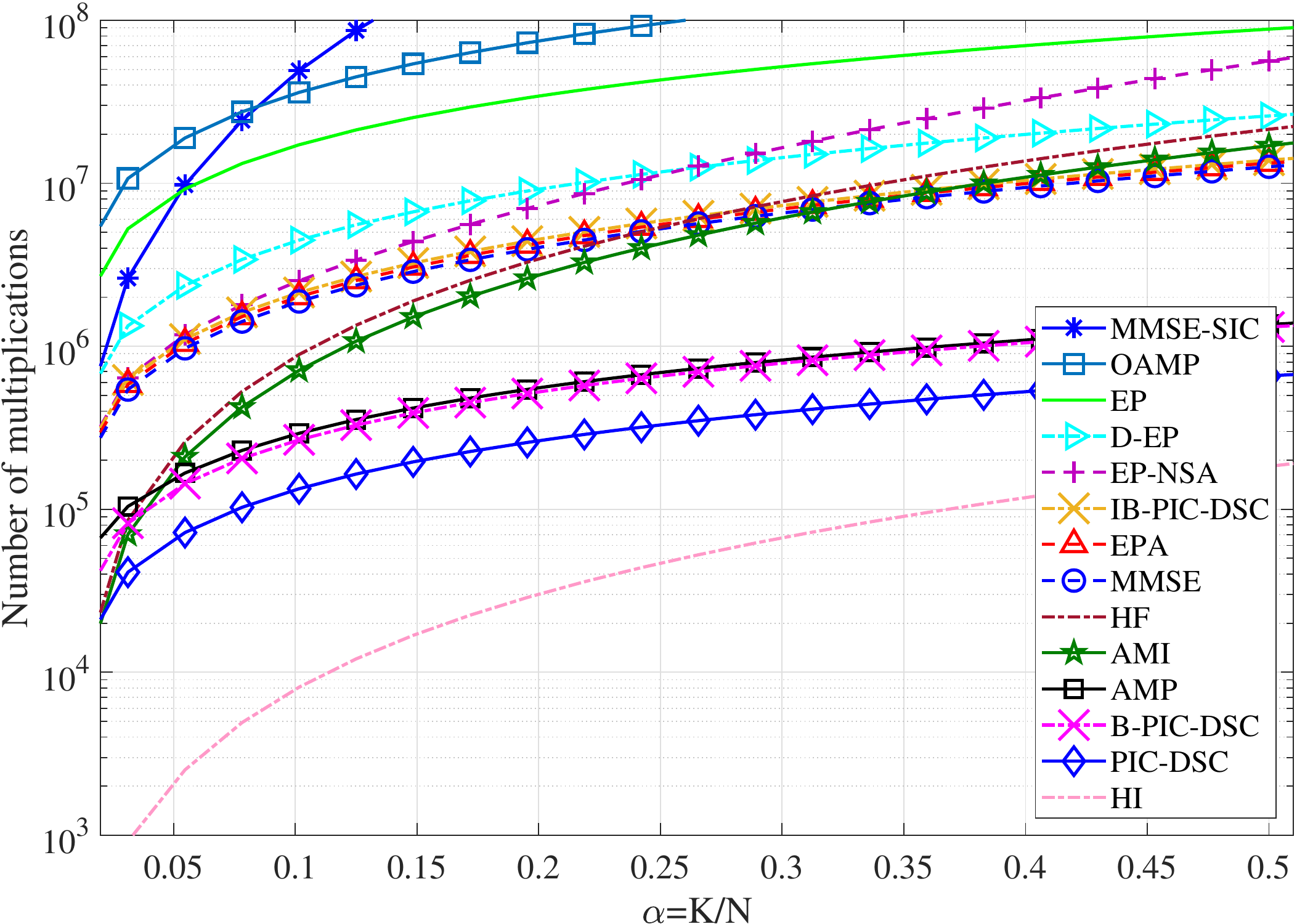}}
  \caption{The number of multiplications needed by the HI \cite{2020ZDan_ICICN_HybridMatrxInv}, HF \cite{2020Chataut_DCOSS_HFADMM}, AMI \cite{2019Albrem_WCNC_AppMatrxInv},  PIC-DSC \cite{PIC2012}, MMSE \cite{LMMSE}, MMSE-SIC \cite{MMSE_SIC2}, AMP \cite{LAMA_Paper}, OAMP \cite{Ma-17ACCESS}, D-EP \cite{A.Kosasih}, EP-NSA \cite{EP-NSA2018}, EPA \cite{EP2019}, EP \cite{Jespedes-TCOM14},  B-PIC-DSC, and IB-PIC-DSC detectors}
  \label{F_C}
\end{figure}

The variance of the BSE module  in the $t$-th iteration is expressed as
\begin{equation}\label{BSE_Analytical}
V_k^{(t)}=  \Ex  \left[ \left| x_k  - \Ex\left[x_k \Big| x_{{\rm PIC},k}^{(t)} ,\Sigma_k^{(t)} \right] \right|^{2} \right].
\end{equation}
Although \eqref{BSE_Analytical} is just a restatement of  \eqref{eA1_b02}, we write it to emphasize that $ x_{{\rm PIC},k}^{(t)} $ in \eqref{BSE_Analytical} must follow Assumption 1, expressed in \eqref{Awgn}. We can then iterate the computation of $(v^{(t)},V_k^{(t)})$ in  \eqref{MSE-IB-PIC-DSC} or \eqref{SINR_BSO} and \eqref{BSE_Analytical} until $v^{(t)}$ converges to a certain value,
\begin{equation}\label{Th_conv}
\|v^{(t)}-v^{(t-1)}\|\leq \zeta.
\end{equation}
The BER approximation of the proposed detectors with $4$-QAM modulation can then be obtained by using the relation of the BER and SINR given in \cite{D.Yoon2000} as
\begin{equation}\label{BER_SE}
{\sf BER}_{\rm ap }= \frac{1}{2}  {\rm erfc} \left( \sqrt{\frac{1}{v^{(T)}}}\right),
\end{equation}
where $T$ is the total number of iterations and erfc is the complementary error function. Note that in case of $M \neq 4$, an appropriate ${\sf BER}_{\rm ap }$ expression can be found by substituting the obtained SINR $1/v^{(T)}$ into the BER expression provided in  \cite{D.Yoon2000}. The pseudo-code of the BER approximation of the proposed detectors is shown in Algorithm \ref{A3}. 

\begin{table*}\small
\label{table1}
\centering
\caption{Number of multiplications comparison}
 \begin{threeparttable}
 \begin{tabular}{|l|c|r|} \hline 
 Detector & Number of multiplications\tnote{*}\\ \hline \hline
AMI with Gauss Seidel \cite{2019Albrem_WCNC_AppMatrxInv} & 	$(4N + 4 T_{\rm AMI} - 2)K^2 + 2(N - 2 T_{\rm AMI} +1)K$		\\ \hline
HF-ADMM \cite{2020Chataut_DCOSS_HFADMM} & 		$2NK^2 + (N + 1)K +(NK^2 + 9 K^2)T_{\rm HF-ADMM}$		\\ \hline
HI \cite{2020ZDan_ICICN_HybridMatrxInv} & 	$5K^2 + K - 6 + (2K^2 + 8K +6)T_{\rm HI}$				\\ \hline
MMSE \cite{LMMSE} & 		$(N+1)K^2 +N^2K+NK  $		\\ \hline
MMSE-SIC  \cite{MMSE_SIC2} & $\sum_{k=0}^K ((N+1)k^2 + N^2k+Nk  )$\\ \hline
PIC-DSC\cite{PIC2012}& $ 4 (N+1)KT_{\rm PIC}$\\ \hline
AMP \cite{LAMA_Paper} & $(4NK + 8N +6K + 4 M K) T_{\rm AMP}$\\ \hline
OAMP \cite{Ma-17ACCESS} & $ (K-1)NK + (2N^2K + NK^2 + 2NK + 12K + 4  M K +8) T_{\rm OAMP}$      \\ \hline
B-PIC-DSC & $  (4NK + 12K + 4 M K) T_{\rm B-PIC-DSC} -(NK+5K) $ \\ \hline 
IB-PIC-DSC & $N^2K + NK^2 -4K -2NK + (4NK + 12K + 4 M K) T_{\rm IB-PIC-DSC} $\\ \hline
EP-NSA, $T_{\rm NSA}=2$  \cite{EP-NSA2018} &   $ (N+1)K^2 +( N^2+N+1)K  + ((K + 1)2K^2+(4N+ 4 M+14)K  )T_{\rm EP-NSA}$               \\ \hline
EPA  \cite{EP2019} &   $ (N+1)K^2  + (N^2-1)K + (2N + 4M +8)KT_{\rm EPA}  $     \\ \hline
D-EP  \cite{A.Kosasih} & $ ( K^2 + K )N_c  C + (N_c^2  C + KC + 3C  +4 M + 17)KT_{\rm D-EP}   $        \\ \hline
EP  \cite{Jespedes-TCOM14} & $NK^2 + (N-1)K + (N^2 K + K^2 + 19K + 4 MK)T_{\rm EP}$\\ \hline
\end{tabular}
\begin{tablenotes}
\item[*] $T_{\rm Algorithm}>0$ represents the number of iterations for the designated algorithms, $N_c$ is the number of receive antennas in a cluster in D-EP detector, and $C$ denotes the number of clusters used in the D-EP detector.
\end{tablenotes}
\end{threeparttable}
\end{table*}

\begin{Remark}
The BER approximation of the proposed detectors, derived based on the SINR-variance evolution technique, refers to the recursion of $({v}^{(t)},V_k^{(t)})$ in \eqref{SINR_BSO}, \eqref{BSE_Analytical}, and ${\sf BER}_{\rm ap }$ in \eqref{BER_SE}.  The BER approximation can  be used to predict the performance of the proposed detectors with high accuracy as shown in \ref{Sec:app3}. In practice, we can use the BER approximation algorithm to calculate the maximum number of users that can be served simultaneously with a certain BER requirement. This  allows us to optimize the M-MIMO multiplexing gain. As the BER approximation algorithm has the closed-form expressions, it can be run very fast. Therefore, we are able to optimize the M-MIMO multiplexing gain based on the results from the BER approximation in real time. 
\end{Remark}

\section{Complexity Analysis}
\label{sComplAnalysis}

In this section, we analyse the complexity of the proposed B-PIC-DSC and IB-PIC-DSC detectors in terms of the number of multiplications. Algorithm 1 specifies that the B-PIC-DSC detector performs the operations in \eqref{eA1_a02}, \eqref{eA1_a01}-\eqref{DSC_error}, at each iteration. These operations can be implemented with low complexity by reusing results of intermediate computations.  The number of multiplications needed in the first iteration is $3NK +7K +4MK$. From the second iteration onwards, the B-PIC-DSC detector performs $4NK + 12K + 4 MK $ multiplications in each iteration. Therefore, the total number of multiplications needed by the B-PIC-DSC detector is   $  (4NK + 12K + 4 M K) T - (NK+5K) $.  It can be seen from Algorithm 1 that $T>0$. The  IB-PIC-DSC detector  performs identical computations to that of the B-PIC-DSC detector  except, in the first iteration, replacing $NK$ multiplications, needed to calculate the diagonal of channel Gram matrix, by $N^2K + NK^2+K$ multiplications, required to calculate the MMSE matrix inversion operation.  Therefore, the overall number of multiplications needed by the IB-PIC-DSC detector is $N^2K + NK^2 -4K -2NK + (4NK + 12K + 4 M K) T$.

Table I shows the complexity of the proposed detectors  in comparison with the state-of-the-art detectors. The AMI \cite{2019Albrem_WCNC_AppMatrxInv}, HF-ADMM \cite{2020Chataut_DCOSS_HFADMM}, and HI  \cite{2020ZDan_ICICN_HybridMatrxInv} detectors are low complexity detectors used to approximate the MMSE detector \cite{LMMSE}. The PIC-DSC  detector \cite{PIC2012} is a low complexity detector that has a similar performance to that of the MMSE detector \cite{Branka_PIC_book}. The MMSE-SIC detector   \cite{MMSE_SIC2}  can improve the performance of the MMSE detector at the cost of higher complexity. The EP-NSA, D-EP, and EPA detectors are recent low complexity detectors that can approach the performance of the EP detector \cite{Jespedes-TCOM14}. The AMP \cite{LAMA_Paper} and OAMP \cite{Ma-17ACCESS} detectors are state-of-the-art Bayesian detectors. We can see from the table that the B-PIC-DSC detector has a  lower  complexity compared to the other detectors, e.g. the MMSE detector. This is because the B-PIC-DSC does not need to perform the matrix inversion operation.  The complexity of the M-MIMO detectors in terms of the number of multiplications is also illustrated by Fig. \ref{F_C}. Detectors based on the same principle, e.g. MMSE approximation \cite{2019Albrem_WCNC_AppMatrxInv,2020Chataut_DCOSS_HFADMM,2020ZDan_ICICN_HybridMatrxInv,PIC2012}, need a similar number of iterations, therefore we set the number of iterations for the AMI \cite{2019Albrem_WCNC_AppMatrxInv,2020Chataut_DCOSS_HFADMM}, HF-ADMM \cite{2020Chataut_DCOSS_HFADMM}, HI \cite{2020ZDan_ICICN_HybridMatrxInv}, and PIC \cite{PIC2012} detectors to five, while the other detectors perform ten iterations. Both figures show that the MMSE-SIC, OAMP, and EP detectors have a very high complexity, decreased by several low complexity detectors, i.e. EP-NSA, D-EP, and EPA   detectors, by reducing the matrix inversion operations. The IB-PIC-DSC has a similar complexity compared to the EPA and MMSE detectors, as they need to perform the matrix inverse operation once. The AMP detector exhibits a comparable complexity to the B-PIC-DSC detector, however its BER performance is much worse than the B-PIC-DSC detector, as shown in Section \ref{Performance_eval}.

\section{Polar-Coded M-MIMO Receiver }
\label{sPolar}

\begin{figure*}
\centering
{\includegraphics[scale=0.5]{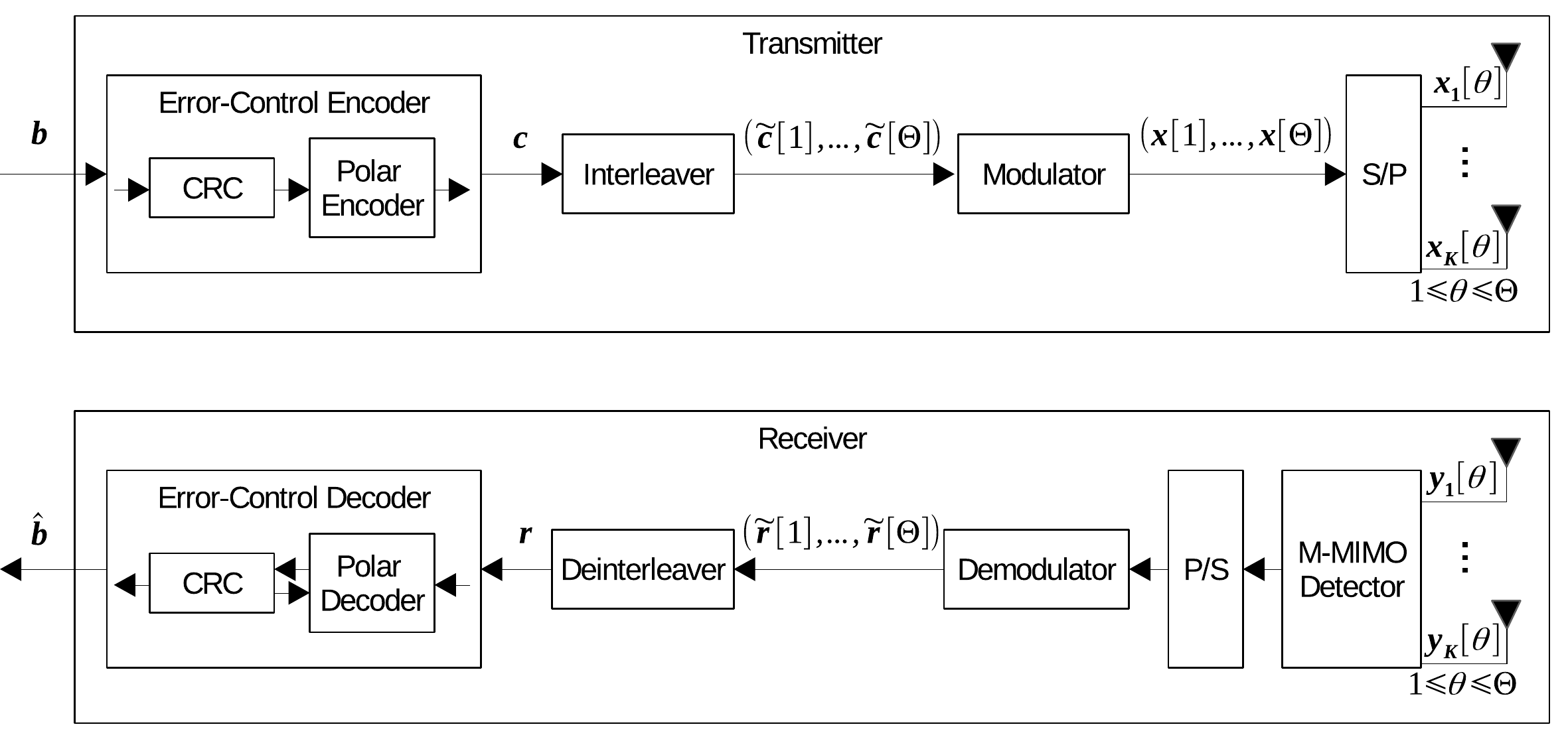}}
\caption{A polar-coded M-MIMO system}
\label{F1}
\end{figure*}
In this section, we propose a high performance M-MIMO receiver to support short packet transmissions for URLLC services, where polar codes could be used \cite{Sharma2019PolarCA}. 
We integrate the proposed B-PIC-DSC detector with a low-complexity polar code decoder. 
Note that to deploy the IB-PIC-DSC detector in a polar coded system, we only need to set the initial estimates of the transmitted symbols as in \eqref{improve_first_iter}. 

\subsection{System Model}

A block-diagram of a 
polar-coded M-MIMO system is shown in Fig. \ref{F1}. At the transmitter side, an error-control 
encoder produces a binary codeword $\qc$ of length $\eta=m\cdot K\cdot \Theta$ for a given binary information vector $\qb$ of length $\eta\cdot R$, where $m=\log_2 M$, $R$ is the code rate and $\Theta$ is an integer parameter. Bits of the codeword $\qc$ are shuffled by an interleaver and split into $\Theta$ blocks  of $m\cdot K$ bits to produce a sequence $(\widetilde\qc[1],\dots,\widetilde\qc[\Theta])$. Note that a random interleaver was used to obtain numerical results for Section \ref{sNumericPolar}. Then, a modulator maps groups of $m$ bits of the interleaved codeword $(\widetilde\qc[1],\dots,\widetilde\qc[\Theta])$ to the symbols of the signal constellation. 
The resulting sequence $(\qx[1],\dots,\qx[\Theta])$ is parallelized into $K$ 
streams $(\qx_k[1],\dots,\qx_k[\Theta])$ by a serial-to-parallel converter (S/P), $1\leq k\leq K$. Each stream $(\qx_k[1],\dots,\qx_k[\Theta])$ is further transmitted to the receiver by the $k$-th antenna during $\Theta$ time slots, $1\leq k\leq K$. Transmission through Rayleigh fading channel with M-QAM modulation is considered in this paper.
 
A received signal block $\qy[\theta]$ corresponding to transmitted block $\qx[\theta]$ is described by  \eqref{eII_1}, where the MIMO channel in the $\theta$-th time slot is characterized by the $N\times K$ matrix $ \qH[\theta] $. The signal blocks $\qy[\theta]$, $1\leq \theta\leq \Theta$, are independently processed by the B-PIC-DSC detector, which is illustrated in Fig. \ref{F0}. For each $\theta$,  the B-PIC-DSC detector  iteratively computes \eqref{eA1_a02}, \eqref{eA1_a01}-\eqref{assign} to yield $x^{(T)}_{{\rm PIC},k}[\theta]$ and its  variance ${\Sigma}^{(T)}_k[\theta]$, which are further used by the demodulator to compute LLR for the $q$-th bit of the $k$-th user symbol transmitted in the time slot $\theta$ according to
\begin{equation}\label{LLR_C}
\widetilde{r}_{(k-1)\cdot m+q}[\theta]=\\
{\sf log}\frac{\sum_{x_k[\theta] \in \Omega_q^{(0)}}\calN\left(x_k[\theta],x^{(T)}_{{\rm PIC},k}[\theta];{\Sigma}^{(T)}_k[\theta]\right)}{\sum_{x_k[\theta] \in \Omega_q^{(1)}}\calN\left(x_k[\theta],x^{(T)}_{{\rm PIC},k}[\theta];{\Sigma}^{(T)}_k[\theta]\right)},
\end{equation}
where $1\leq k\leq K$, $1\leq q\leq m$, and $\Omega_q^{(0)}$ and $\Omega_q^{(1)}$ are the subsets of $\Omega$  consisting of the constellation points corresponding to user's symbols with the q-th bit equal to 0 and 1, respectively.
The LLRs $\widetilde{\qr}[\theta]=(\widetilde{r}_1[\theta],\dots,\widetilde{r}_{K\cdot m}[\theta])$, $1\leq \theta\leq \Theta$, are combined into a single sequence and deinterleaved. The resulting sequence $\qr$ consisting of $m\cdot K\cdot \Theta$ LLRs is sent to a polar code decoder to compute an estimate $\hat\qb$ of the original information vector $\qb$. 

\subsection{Polar Codes}

A $(\eta=2^\mu,\kappa)$ polar code \cite{arikan2009channel} is a linear block code generated by $\kappa$ rows of the matrix $B_\eta\cdot G_2^{\otimes \mu}$,
where $G_2=\begin{pmatrix}1&0\\1&1\end{pmatrix}$, $\mu\in\mathbb{N}$,
$\otimes \mu$ denotes $\mu$-times Kronecker product of a matrix with itself and $B_\eta$ is $\eta\times\eta$ bit reversal permutation matrix. 
Any codeword of a polar code can be represented as $\qc=\qu\cdot B_\eta\cdot G_2^{\otimes \mu}$,
where $\qu=(u_1,\dots,u_\eta)$ is an input sequence, such that $u_i=0$, $i\in\mathcal{F}$,
where $\mathcal{F}\subset\{1,\dots,\eta\}$
is the set of $\eta-\kappa$ indices of frozen bits.
The remaining $\kappa$ elements of $\qu$ are set to the information bits.

It was shown in \cite{tal2011list} that the error-correction capability of polar codes can be substantially improved by concatenating them with cyclic redundancy check (CRC) codes as depicted in Fig. \ref{F1}. 
Polar codes have been adopted by the 3rd generation partnership project (3GPP) for 5G New Radio as a channel coding technique for control information for eMBB scenario \cite{polarNR2018}. CRC codes of length 11 and 24 were selected for the uplink and downlink control information, respectively.

\subsection{Sequential Decoding of Polar Codes and its Integration with B-PIC-DSC Detection}
\label{sPolarSeq}

Let us denote a channel between the polar code encoder and decoder as $W^\eta$: $\{0,1\}^\eta\rightarrow\mathbb{R}^\eta$. 
Given a polar code $C$ and a received vector $\qr$, the decoding problem consists in finding $\widehat \qc=\arg \max_{\qc\in C} W^{\eta}(\qc|\qr)$. This problem is equivalent to finding $\widehat \qu=\arg \max_{\qu} W^{\eta}(\qu|\qr)$ since $\qc=\qu\cdot B_\eta\cdot G_2^{\otimes \mu}$, where maximization is performed over the set of vectors $\qu\in\{0,1\}^\eta$ satisfying constraints imposed by $\mathcal{F}$. Recursive structure of polar codes enables low-complexity decoding using the SC algorithm \cite{arikan2009channel} and its list/stack variations such as the sequential decoding algorithm \cite{miloslavskaya2014sequential}. 
These algorithms keep one or several of the most probable paths $u_1^{i} \triangleq(u_1,\dots,u_i)\in\{0,1\}^i$ within the code tree and sequentially make decisions on input bits $u_i$ for $i=1,\dots,\eta$, where each path is associated with the corresponding score characterizing its probability. 
Similarly to the SCS \cite{niu2012crcaided}, the sequential algorithm   
keeps the paths  in a stack (priority queue). 
At each iteration, the decoder selects for extension path $u_1^{i}$ with 
the largest score,
and 
computes the score for path $(u_1^{i},0)$ and, if $(i+1)\notin\mathcal{F}$, also for path $(u_1^{i},1)$, then puts the path(s) into the stack. 
Once the decoder constructs $L$ paths of length $i$, 
all paths shorter than $i$ are eliminated in order to keep the size of the stack limited. Parameter $L$ is called the list size.
Decoding terminates as soon a path of length $\eta$ appears at the top of the stack, or the stack becomes empty.  
Hence, the worst case complexity of such decoding is given by $O(L\cdot\eta\cdot\log \eta)$. Average decoding complexity depends on how path scores are defined.

The sequential decoding algorithm provides a significant complexity reduction compared to the SCS and SCL with a negligible performance degradation as shown in \cite{miloslavskaya2014sequential,Trifonov2018ASF}. The complexity reduction is achieved by redefining the score function: simplifying recursive calculation and introducing a bias function to estimate the conditional probability of the most likely
codeword of a polar code. 
According to the sequential decoding algorithm, a path $u_0^i$ is associated with the following score 
\begin{equation}
\hat T(u_1^i,\qr) = R(u_1^i,\qr) \hat\Omega(i),
\label{eqHatT}
\end{equation}
where 
\begin{equation}
R(u_1^i,\qr)=\max_{u_{i+1}^\eta}P(u_1^{\eta}|\qr), \label{eqR}
\end{equation}
\begin{equation}
\hat\Omega(i)=\prod_{j\in\mathcal{F},j>i}(1-P_j),
\label{eqHatOmega}
\end{equation}
where $P_j$ is  the $j$-th subchannel error probability, provided that exact values of all previous bits $u_{j'}$, $j'<j$, are available. 

Computation of probability $R(u_1^i,\qr)$ for code of length $\eta$ reduces to computation of two probabilities for codes of length $\eta/2$, i.e. 
$$R(u_1^{2i-1},r_1^\eta)=\max_{u_{2i}\in\{0,1\}} R\left(u_{1,o}^{2i}\oplus u_{1,e}^{2i},r_1^{\eta/2}\right)\cdot R\left(u_{1,e}^{2i},r_{\eta/2+1}^{\eta}\right),$$
$$R(u_1^{2i},r_1^{\eta})=R\left(u_{1,o}^{2i}\oplus u_{1,e}^{2i},r_1^{\eta/2}\right)\cdot 
R\left(u_{1,e}^{2i},r_{\eta/2+1}^{\eta}\right),$$
where $u_{1,o}^i$ and $u_{1,e}^i$ are subsequences of $u_1^i$ consisting of elements with odd and even indices, respectively. 
The initial values for these recursive expressions are defined by $\qr$. 
Note that the computations can be efficiently performed in log-domain using only addition and comparison operations \cite{miloslavskaya2014sequential,Trifonov2018ASF}.

The bias function 
$\hat\Omega(i)$ is equal to the mean value of probability that frozen symbols in the remaining part of input sequence $u_{i+1}^{\eta}$ are equal to $\mathbf{0}$. It depends only on $\eta$, $\mathcal{F}$ (i.e. the code being considered), channel properties and phase $i$.
This approach enables one to compare paths $u_1^i$ of different lengths, and prevent the decoder from switching frequently between different paths.
For any given channel, probabilities $P_j$ for the bias function $\hat\Omega (i)$ can be pre-computed offline using  density evolution \cite{Mori2009PerformanceAC,tal2011how}.
A low-complexity alternative based on Gaussian approximation of density evolution 
was presented in \cite{trifonov2012efficient} for a binary AWGN channel. In case of \cite{trifonov2012efficient}, only the noise and interference variance is needed to compute probabilities $P_j$ for $1\leq j\leq\eta$. Min-sum density evolution \cite{Kern2014ANC} provides a tradeoff between high accuracy and low complexity. 
An improved score function based on min-sum density evolution was proposed for sequential decoding in \cite{Trifonov2018ASF}.  

It can be seen that $W^\eta(\qr|\qc)=\widetilde W^\eta\Big((\widetilde\qr[1],\dots,\widetilde\qr[\Theta])|(\widetilde\qc[1],\dots,\widetilde\qc[\Theta])\Big)$, where $\widetilde W^\eta$: $\{0,1\}^\eta\rightarrow\mathbb{R}^\eta$ is a channel between the interleaver and the deinterleaver. Recall that 
the blocks $\widetilde\qc[\theta]$, $1\leq\theta\leq\Theta$,  are transmitted independently through a  memoryless 
channel and that $\eta=\Theta\cdot m\cdot K$. Thus, channel $\widetilde W^\eta$ can be decomposed into $\Theta$ independent parallel channels $\widetilde W^{m\cdot K}$, more specifically, $\widetilde W^\eta\Big((\widetilde\qr[1],\dots,\widetilde\qr[\Theta])|(\widetilde\qc[1],\dots,\widetilde\qc[\Theta])\Big)=\prod_{\theta=1}^\Theta \widetilde W^{m\cdot K}(\widetilde{\qr}[\theta]|\widetilde\qc[\theta])$. 
Since we consider an M-MIMO scenario in which the number of transmit and receive antennas is large, we can employ an approximation $\widetilde W^{m\cdot K}(\widetilde{\qr}[\theta]|\widetilde\qc[\theta]) \approx \prod_1^K p(x^{(T)}_{{\rm PIC},k}[\theta]|x_k[\theta])$,  
 where pdf $p(x^{(T)}_{{\rm PIC},k}[\theta]|x_k[\theta])$ is characterized by Assumption \ref{assump:x}. From \eqref{Awgn}, we  obtain $p(x^{(T)}_{{\rm PIC},k}[\theta]|x_k[\theta]) = {\cal{N}} \left( x^{(T)}_{{\rm PIC},k}[\theta], x_k[\theta] ; v^{(T)}   \right)$,  where $v^{(T)}$ is given in \eqref{SINR_BSO} and its derivation can be found in Appendix \ref{Ap.1}. The validation of this approximation is equivalent to the validation of the BER approximation of the proposed detector given in Section \ref{Sec:app3} as both approximations are based on Assumption  \ref{assump:x}. The obtained noise and interference variance $v^{(T)}$ can be employed to compute the probabilities $P_j$ as described in \cite{trifonov2012efficient,Kern2014ANC}. The probabilities $P_j$ are further substituted into the bias function 
 for the sequential decoder.

\begin{figure}
  \centering
  \includegraphics[scale=0.34]{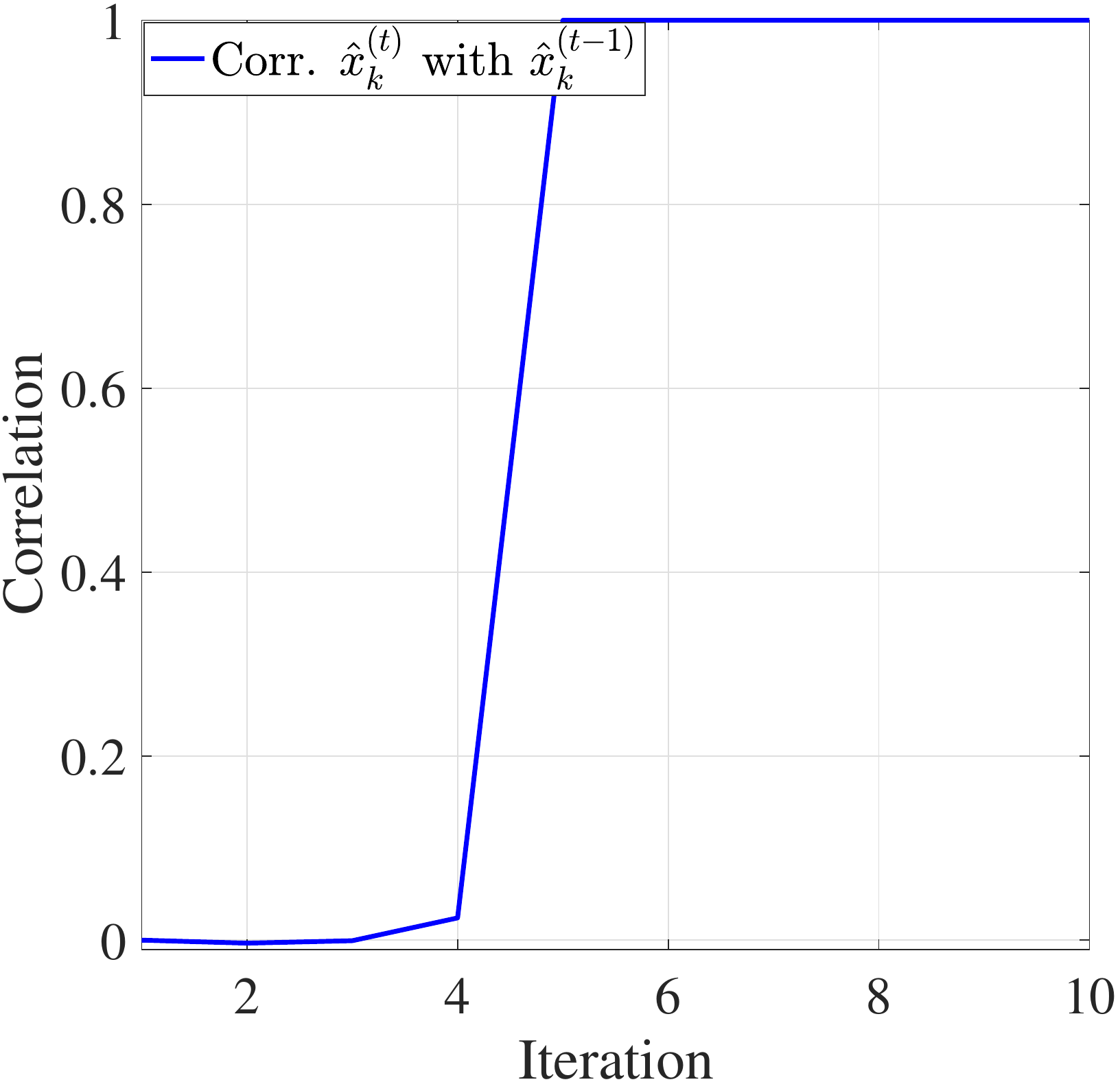}
  \caption{The correlation (corr.) between $\hat{x}_k^{(t)}$ and $\hat{x}_k^{(t-1)}$ in B-PIC-DSC detector with $N=144$, $K=48$, and SNR = $10$ dB}
  \label{F11}
\end{figure}

\section{Numerical Results}\label{Numerical}

In this section, we evaluate the performance of the proposed detectors and polar coded M-MIMO receivers. First, we evaluate the correlation between $\hat{x}_k^{(t)}$ and $\hat{x}_k^{(t-1)}$ as a function of the number of iterations $t$ for the proposed B-PIC-DSC detector.  We then analyse the ratio of transmit-to-receive antennas $\alpha=K/N$ for uncoded transmission that ensures a target BER of $10^{-5}$ required for  $5$G systems \cite{3GPP_Phy}. We also verify the accuracy of our BER approximation $\sf BER_{\rm ap}$ defined by equation  \eqref{BER_SE}. We then investigate the convergence behaviour of the proposed detectors by plotting the number of iterations needed versus $\alpha$.  Finally, we evaluate the performance of our proposed B-PIC-DSC detector, IB-PIC-DSC detector, and polar coded M-MIMO receiver. The simulation results are obtained by averaging over 1,000,000 channel realizations wherein the detectors are implemented with the maximum number of iterations $T_{\rm max} = 10$ and parameter $\zeta = 10^{-4}$.  The $4$-QAM constellation is  used to transmit data.  

\subsection{The correlation between $\hat{x}^{(t)}_k$ and $\hat{x}^{(t-1)}_k$}\label{Sec:app2}

Fig. \ref{F11} shows the correlation between $\hat{x}^{(t)}_k$ and $\hat{x}^{(t-1)}_k$ for the B-PIC-DSC scheme, where the Pearson correlation coefficient of both variables is   calculated for different realizations of the channel and noise. As illustrated in Fig. \ref{F11}, the correlation between $\hat{x}^{(t)}_k$ and $\hat{x}^{(t-1)}_k$ is very low when the number of iterations is less than $4$. Hence, in the early iterations, $\hat{x}^{(t)}_k$ and $\hat{x}^{(t-1)}_k$ can be used to smoothen transition between iterations using the linear combination in \eqref{DSC}. 

\subsection{Uncoded BER with different values of $\alpha$ }\label{Sec:app3}

\begin{figure}
  \centering
 \includegraphics[scale=0.37]{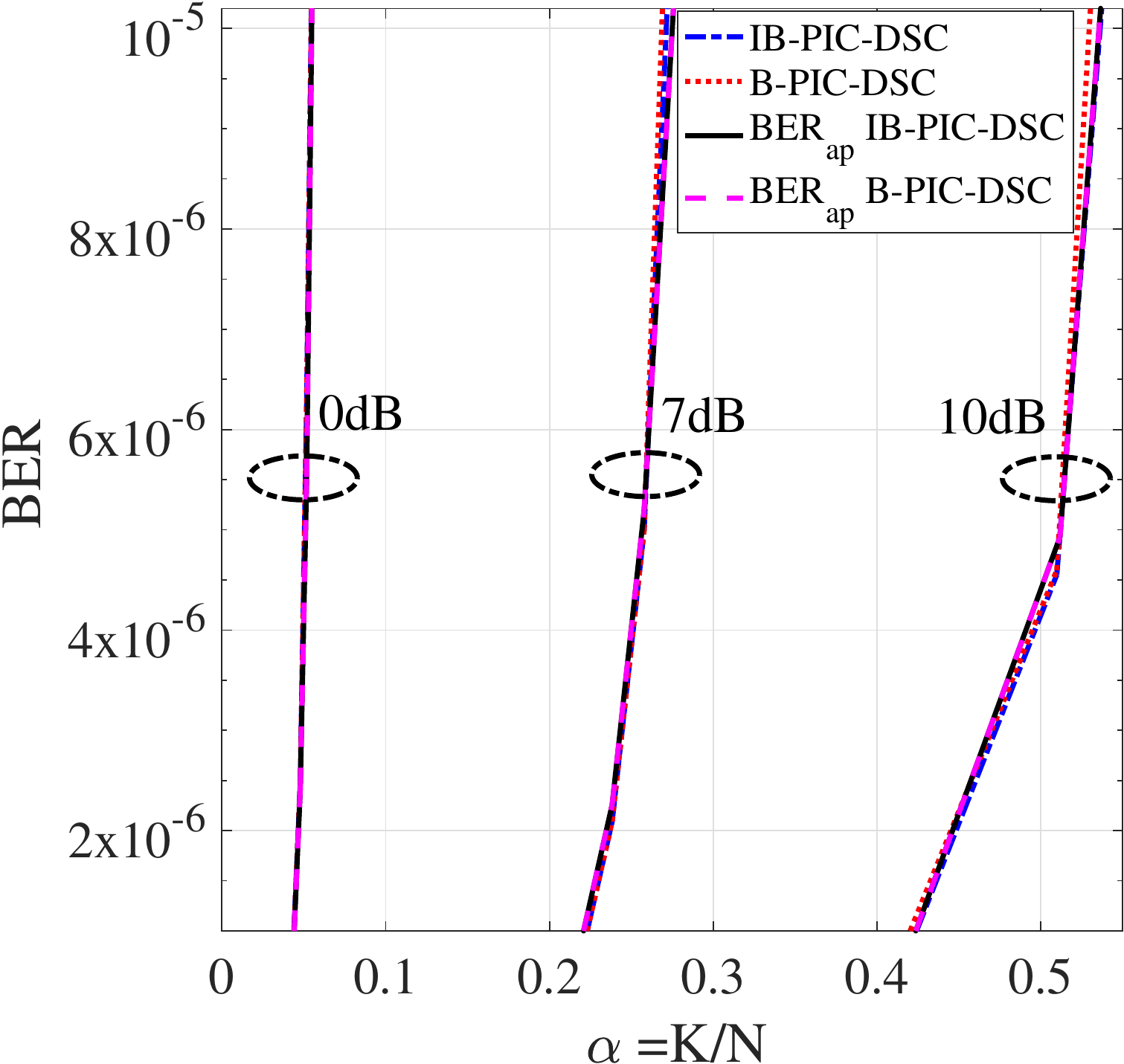}
  \caption{The analysis of $\alpha$ in an uncoded system to achieve BER lower than $10^{-5}$ with $N=1024$}
  \label{F2}
\end{figure}

In this section, we evaluate the BER of the proposed detectors and their BER approximations with different values of $\alpha$ in the case of uncoded transmission. This is used to show how many users can be supported by our receiver to achieve a maximum acceptable BER of $10^{-5}$ required in $5$G system \cite{M-MIMO_URLLC_wireless-access} for SNR $\in \{0,7,10 \}$ dB. The results in Fig. \ref{F2} show that the higher the SNR, the larger the number of users that can be supported by the B-PIC-DSC and IB-PIC-DSC.

\subsection{Convergence Analysis}

\begin{figure}
\centering
\subfloat[N=32 and SNR=10 dB]
{\includegraphics[scale=0.32]{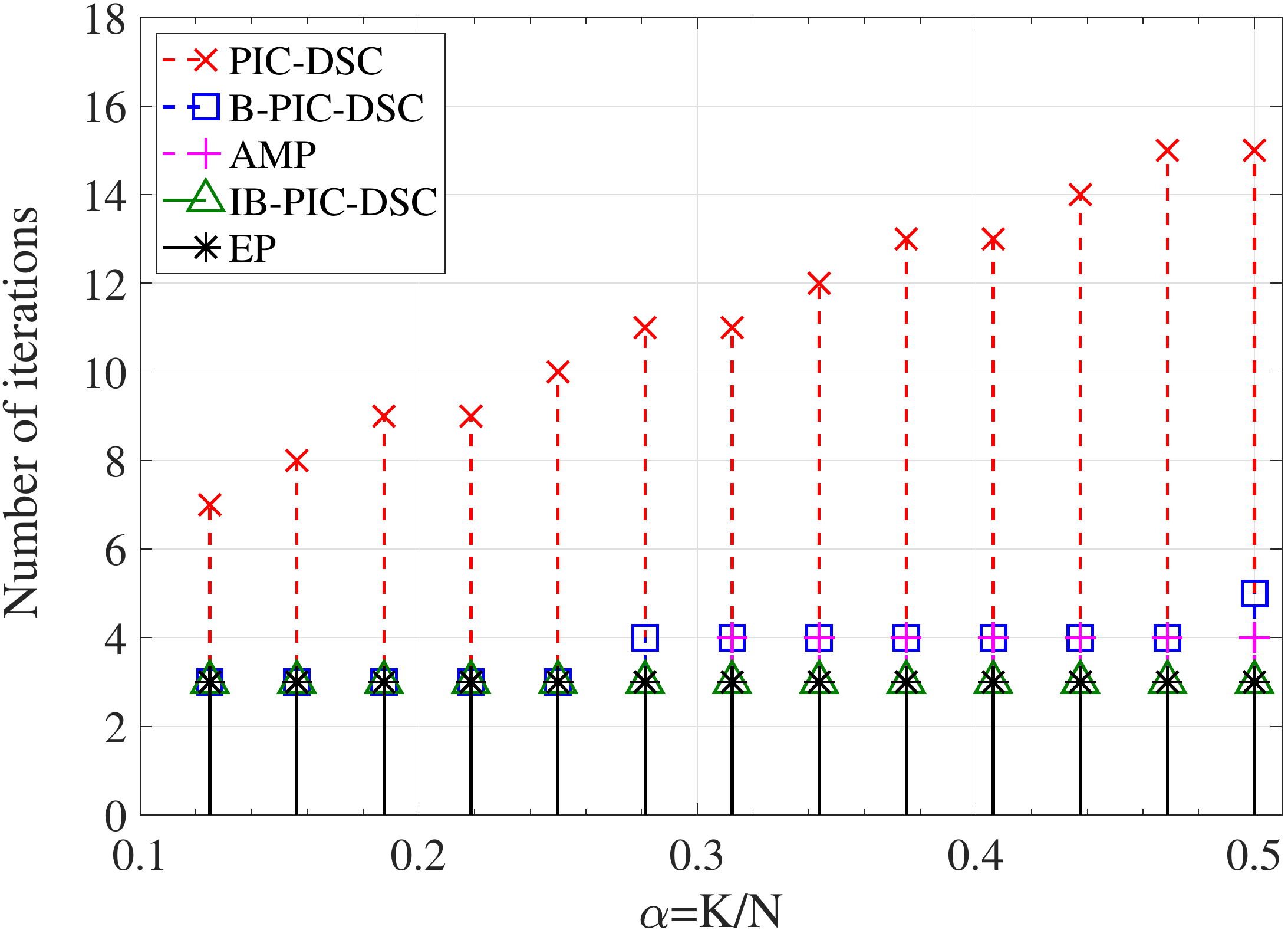}}\hfill
\centering
\subfloat[N=128 and SNR=10 dB]
{\includegraphics[scale=0.32]{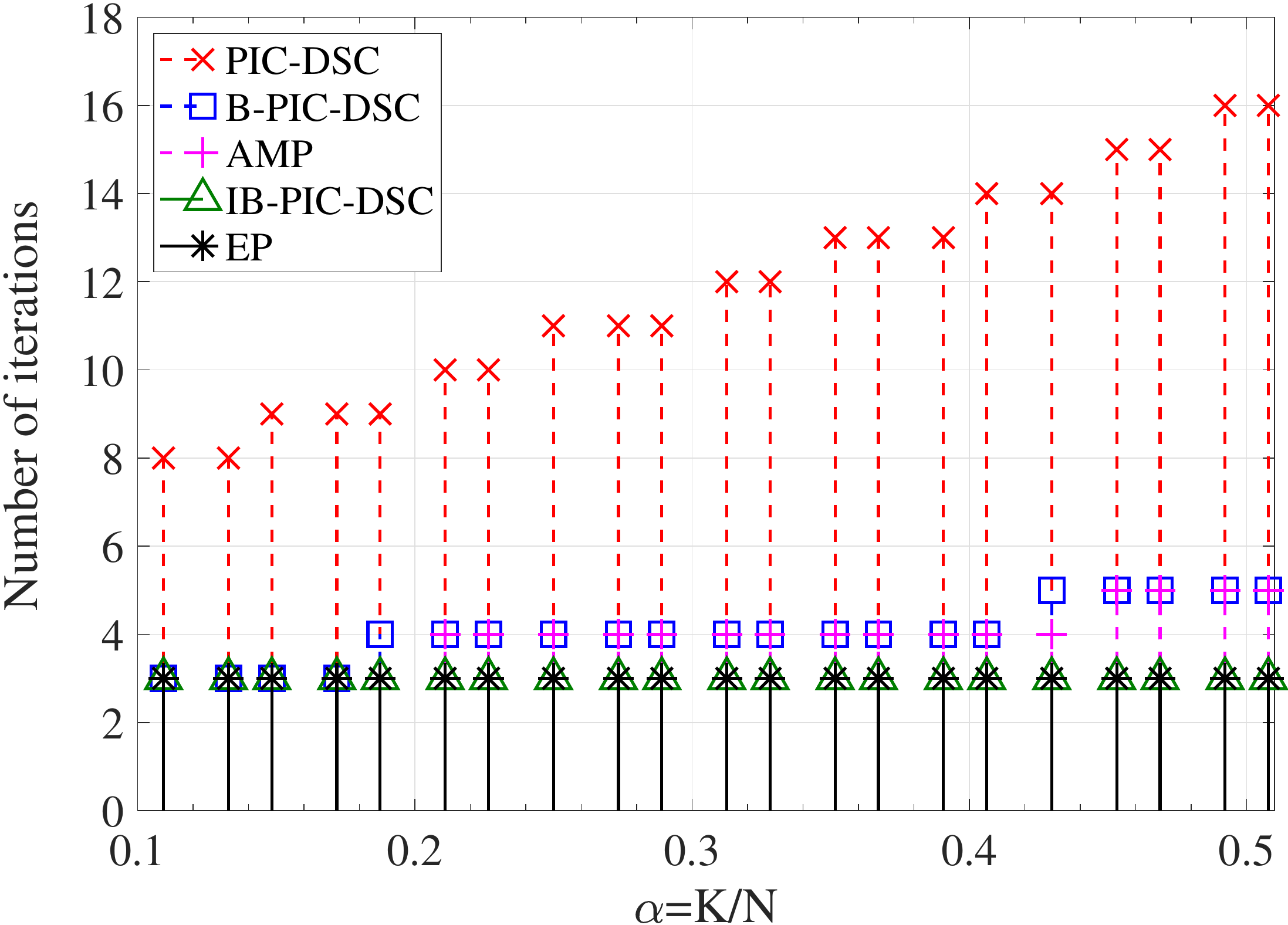}}
  \caption{The convergence behaviour of the PIC-DSC \cite{PIC2012}, B-PIC-DSC, AMP \cite{LAMA_Paper}, IB-PIC-DSC, and EP \cite{Jespedes-TCOM14} detectors in terms of the dependence of the number of iterations on  the ratio of transmit-to-receive antennas  $\alpha$} 
  \label{F4}
\end{figure}

We compare the convergence rate of our proposed B-PIC-DSC and IB-PIC-DSC detectors with the AMP \cite{LAMA_Paper}  and  EP \cite{Jespedes-TCOM14} detectors representing the Bayesian detectors with the lowest and highest complexity, respectively. The comparison is presented by plotting the number of iterations versus $\alpha$. Fig. \ref{F4} shows that all detectors except the PIC-DSC detector need at most $5$ iterations to meet their respective convergence criteria for different values of $\alpha$. 
The B-PIC-DSC and AMP detectors need a slightly higher number of iterations than the EP and IB-PIC-DSC detectors. It can be seen that the convergence rate of the IB-PIC-DSC detector is identical to that of the EP detector. This implies that the lower complexity of the B-PIC-DSC detector comes at the expense of a slightly slower convergence rate. 

\subsection{Uncoded BER versus SNR}\label{Performance_eval}

To illustrate the reliability of the proposed detectors, we compare the uncoded BER achieved by the PIC-DSC \cite{PIC2012}, MMSE \cite{LMMSE}, MMSE-SIC \cite{MMSE_SIC2}, AMP \cite{LAMA_Paper}, OAMP \cite{Ma-17ACCESS}, D-EP \cite{A.Kosasih}, EP-NSA \cite{EP-NSA2018}, EPA \cite{EP2019}, EP \cite{Jespedes-TCOM14}, B-PIC-DSC, IB-PIC-DSC,  and ML \cite{ML}  for different system configurations. The obtained off-line weighting coefficients to smooth output of each iteration for EP, D-EP, EP-NSA, EPA, OAMP, and AMP are $ 0.9, 0.7, 0.5, 0.5, 0.7 $ and $0.2$, respectively. We do not draw the BER of the AMI \cite{2019Albrem_WCNC_AppMatrxInv}, HF-ADMM \cite{2020Chataut_DCOSS_HFADMM}, and HI \cite{2020ZDan_ICICN_HybridMatrxInv} detectors as they are used to approximate the MMSE detector, and therefore their performance is limited by the MMSE performance. Thus, we only plot the BER of the MMSE detector. 

\begin{figure*}
\centering
\subfloat[$N=32$,  $K=8$, $\alpha=0.25$,  $\psi=0$,  $\gamma=0$]
{\includegraphics[scale=0.35]{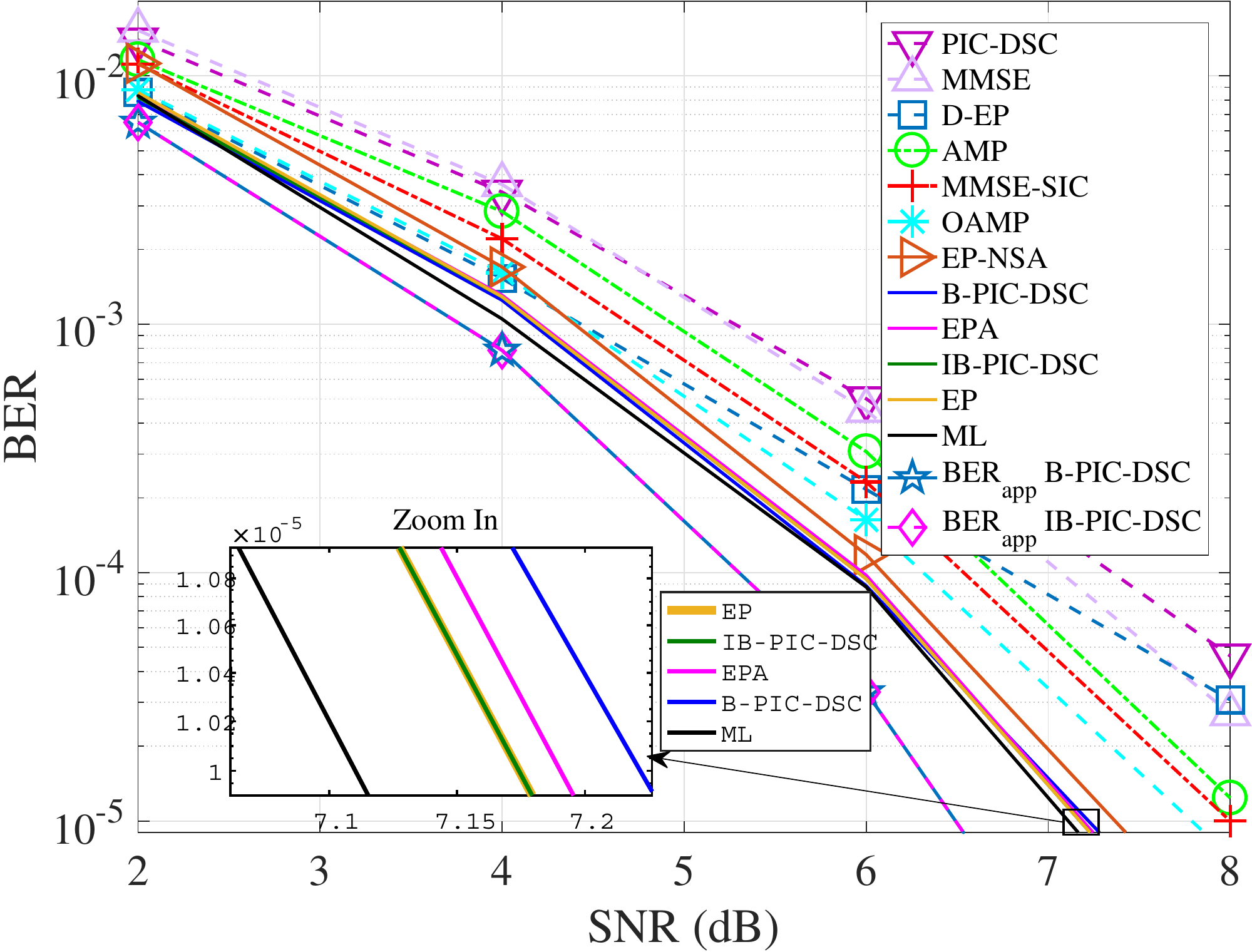}}\hspace{1cm}
\centering
\subfloat[$N=128$,  $K=64$, $\alpha=0.5$, $\psi=0$, $\gamma=0$]
{\includegraphics[scale=0.35]{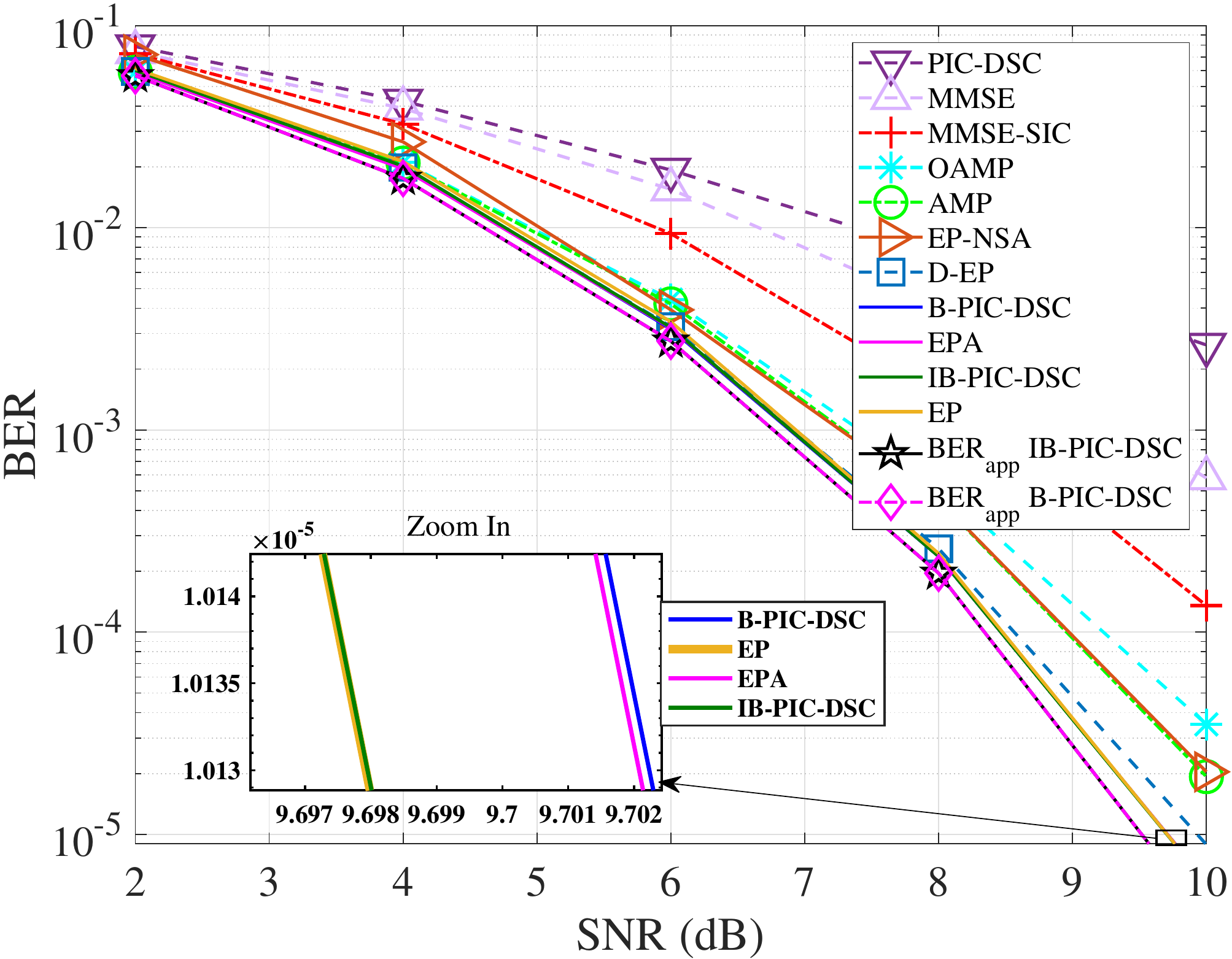}}\hfill
\centering
\subfloat[$N=128$,  $K=4$, $\alpha=0.03125$, $\psi=0.9$, $\gamma=0.2$]
{\includegraphics[scale=0.35]{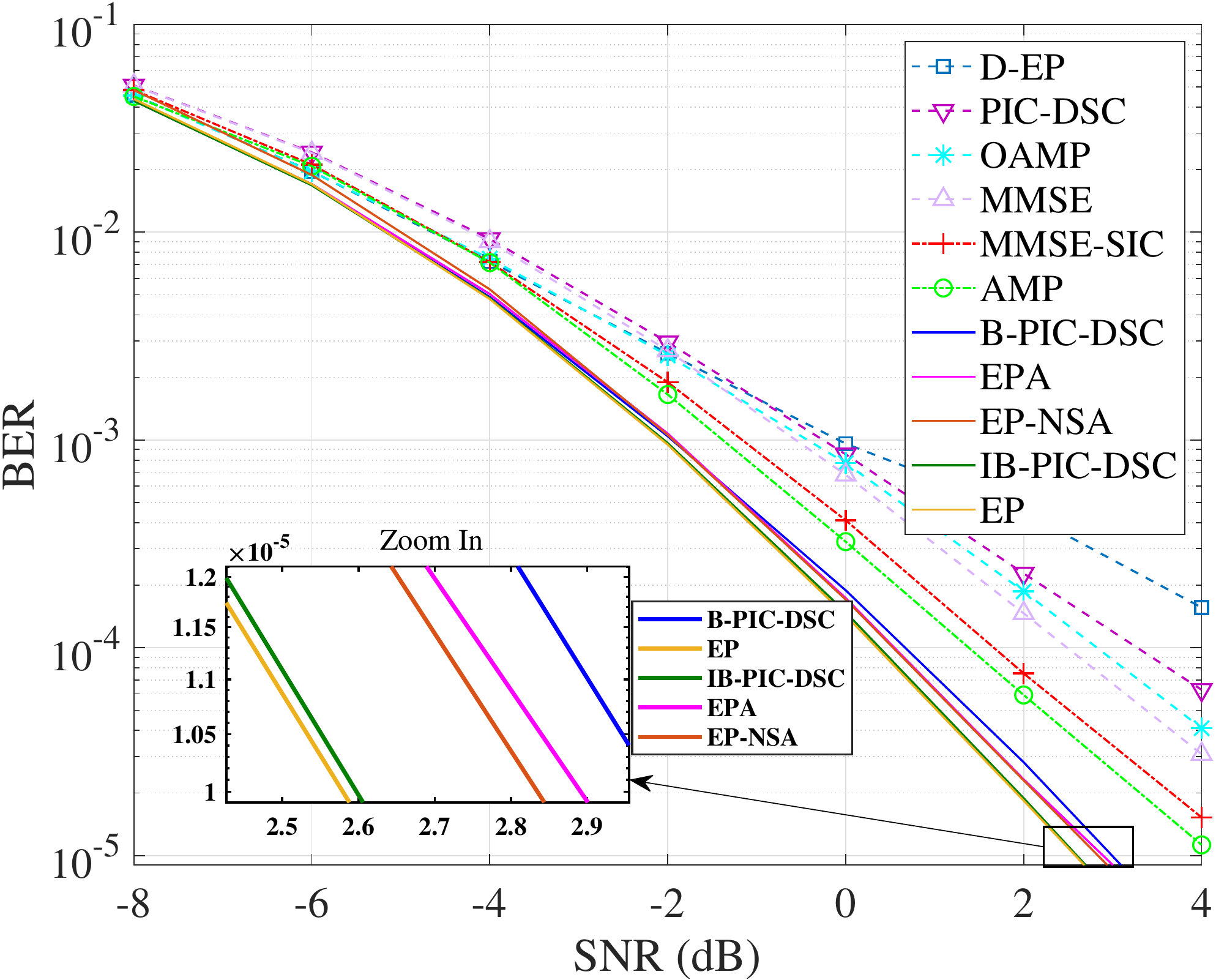}}\hspace{1cm}
\centering
\subfloat[$N=256$,  $K=24$, $\alpha=0.09375$, $\psi=0.7$, $\gamma=0.05$]
{\includegraphics[scale=0.35]{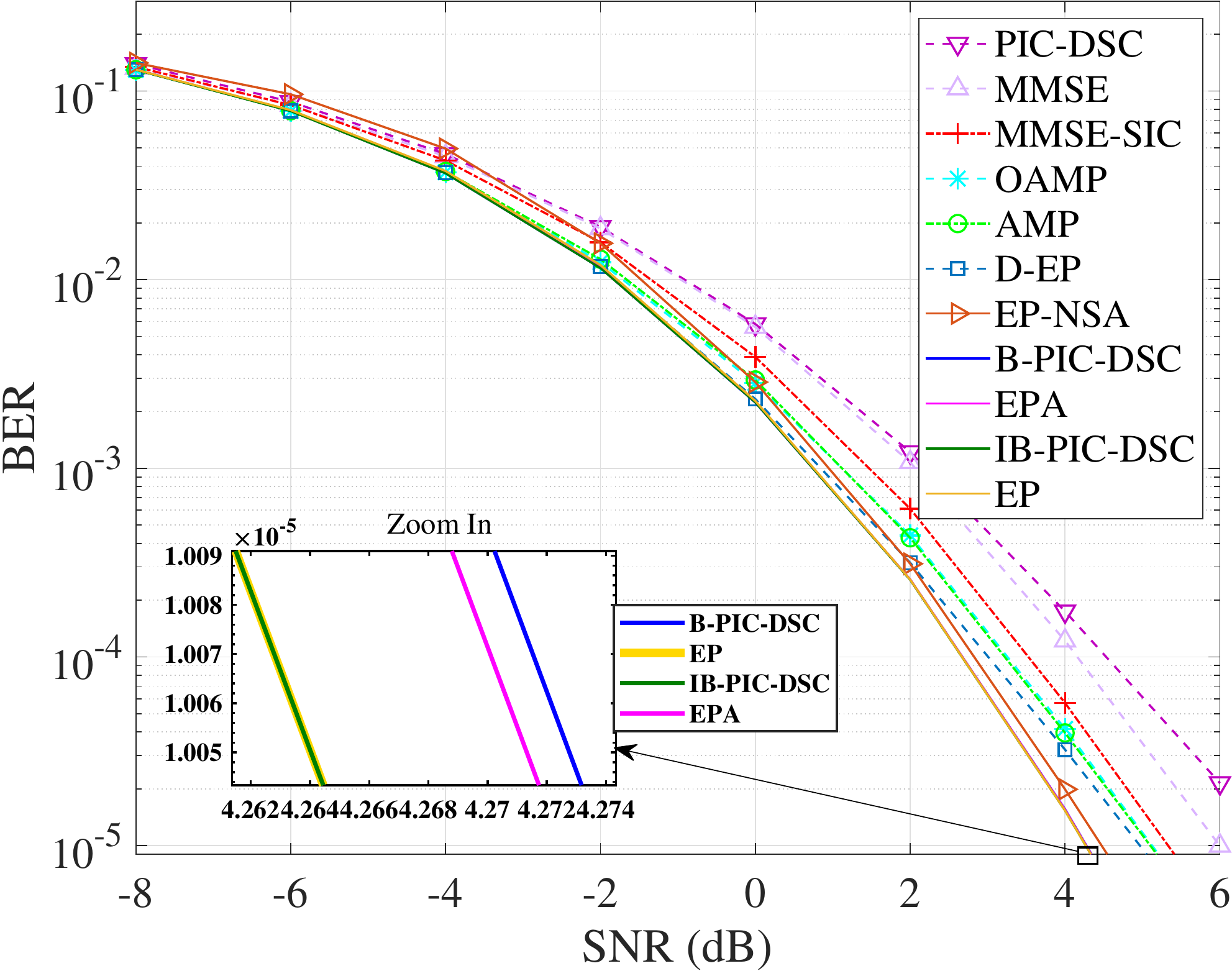}}
\caption{The BER performance of the PIC-DSC \cite{PIC2012}, MMSE \cite{LMMSE}, MMSE-SIC \cite{MMSE_SIC2}, AMP \cite{LAMA_Paper}, OAMP \cite{Ma-17ACCESS}, D-EP \cite{A.Kosasih}, EP-NSA \cite{EP-NSA2018}, EPA \cite{EP2019}, EP \cite{Jespedes-TCOM14}, B-PIC-DSC, IB-PIC-DSC,  and ML \cite{ML} detectors for several system configurations}
\label{F5}
\end{figure*}

The results in Fig. \ref{F5}a and b show that the B-PIC-DSC detector can achieve approximately  $1$~dB performance gain compared to the low complexity AMP and PIC-DSC detectors.  The B-PIC-DSC detector outperforms the other low complexity variants of the EP detector such as the EP-NSA and D-EP detectors. The performance of the proposed detectors is close to the  EP and ML detectors. There is no simulation result for the ML detector in Fig. \ref{F5}b, due to the prohibitively high complexity for $K=64$. 

\begin{figure}
  \centering
  \includegraphics[scale = 0.32]{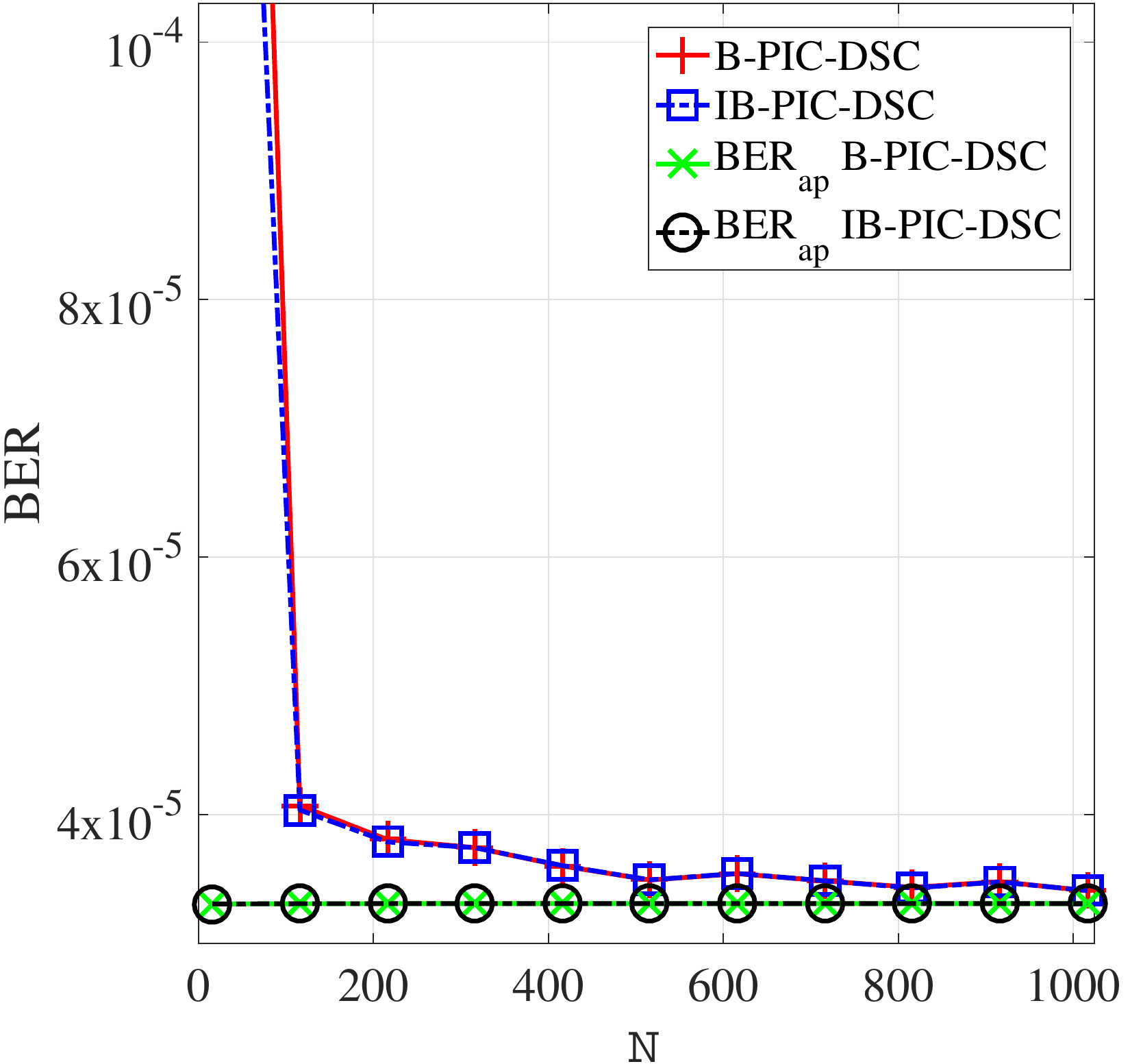}
  \caption{The BER approximation versus the number of receive antennas with $\alpha=0.25$ and SNR = $6$ dB}
  \label{F9a}
\end{figure}

In Fig. \ref{F5}a, the proposed BER approximation shows a significant gap compared to the BER of the B-PIC-DSC and IB-PIC-DSC obtained by the simulation. This is due to a low number of receive antennas in which case the SINR of the BSO module cannot be accurately characterized by \eqref{MSE-IB-PIC-DSC} and \eqref{SINR_BSO}. It can be seen that the BER approximation is much closer to the experimental results for $N=128$ in Fig. \ref{F5}b compared to  $N=32$ in Fig. \ref{F5}a.  To further analyse the accuracy of the BER approximation, we plot Fig. \ref{F9a} which shows that the BER approximation is  accurate for a large number of receive antennas, e.g.  $1024$. Note that the accuracy of the BER approximation for $N=1024$ with several SNR values is illustrated in Fig. \ref{F2}.

To predict the detectors' reliability in real-world settings, we introduce the channel estimation error $\gamma$ and the spatial correlation among receive antennas. As shown in \cite{2007CWANG_TWC_ChErr}, one can model the noisy channel matrix as  $\hat{\qH} = \qH+\gamma \Delta $, 
where $\gamma \in [0,1]$ denotes the magnitude of channel estimation error and each element of vector $\Delta$ follows i.i.d. Gaussian distribution with zero mean and unity variance. It is worth noting that $\gamma \Delta$ is independent with $\qH$. The spatial correlation exists when several receive  antennas are compactly installed such as in M-MIMO systems. We define the spatial correlation as $\qH = \qQ^{\frac{1}{2}} \qH$, where $N\times N$ matrix $\qQ$ characterizes the correlation among the elements of the antenna array, whose $(i, j)$-th entry is set to be $\psi^{|i-j|}$, and  $\psi$ denotes the correlation coefficient. The results are demonstrated  in Fig. \ref{F5}c and d wherein we observe a similar behaviour as in Fig. \ref{F5}a and b. That is, the performance of the B-PIC-DSC detector is comparable to that of the existing  low complexity variants of EP, but achieved with a much lower complexity as explained in Section \ref{sComplAnalysis}, while the IB-PIC-DSC detector can achieve an almost identical performance to the EP detector. Therefore, we conclude that the best trade-off between reliability and processing delay is achieved by the proposed B-PIC-DSC detector. Note that our BER approximation, specified in Algorithm \ref{A3},  is derived by assuming perfect channel state information (CSI) and therefore it is not drawn in Fig. \ref{F5}c and d in which we consider imperfect CSI. Incorporation of the channel estimation errors into the BER approximation will be considered in the future work.

\subsection{Uncoded BER versus Rician factor}\label{Rician}

\begin{figure}
  \centering
 \includegraphics[scale = 0.32]{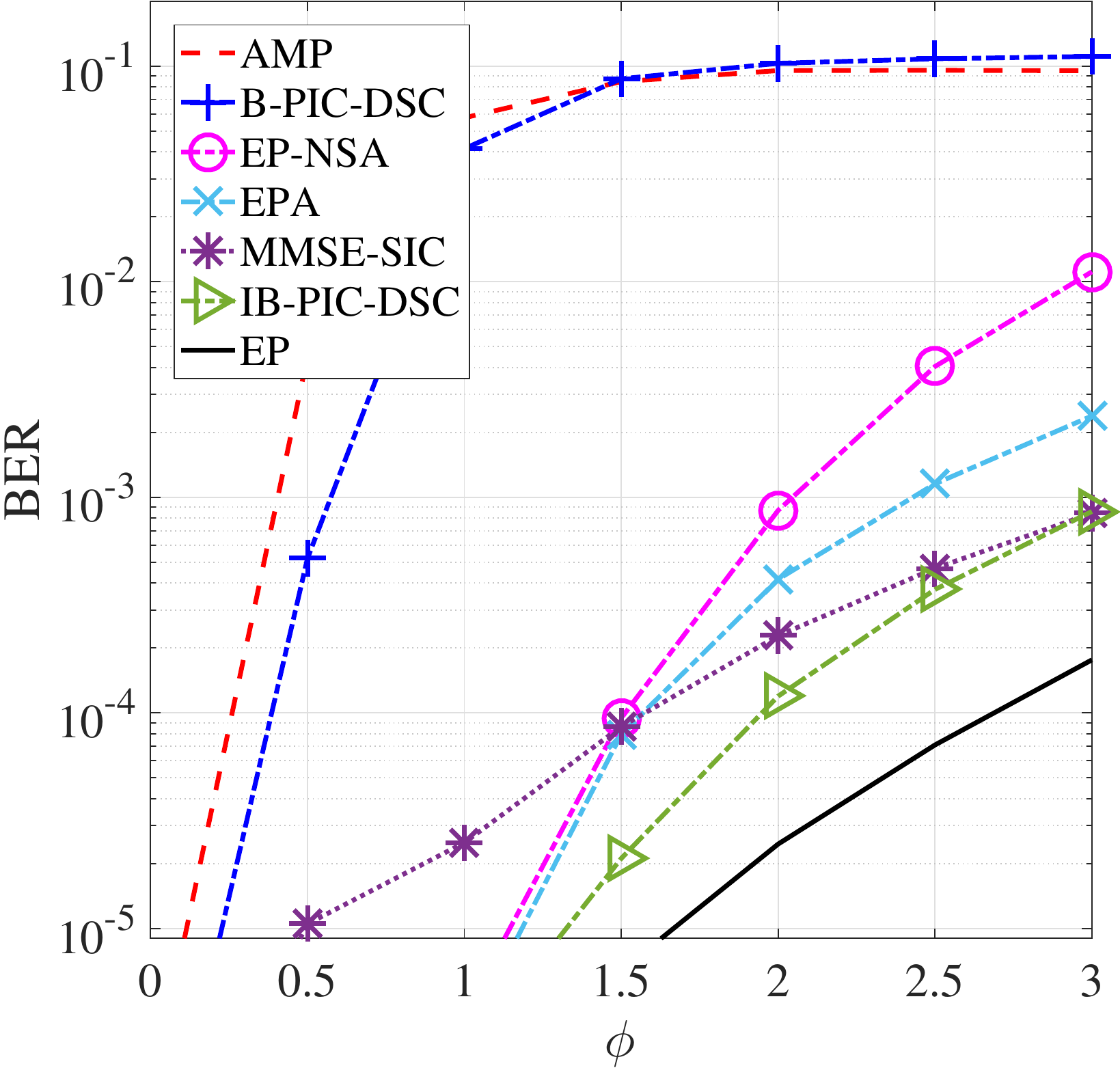}
  \caption{The BER performance evaluation in Rician fading channel with  $N=128$,  $K=8$, $\alpha=0.0625$, $\psi=0$, $\gamma=0$, and SNR = $3$ dB}
  \label{F9b}
\end{figure}

In this section, we evaluate the performance of our proposed detectors, the low complexity AMP detector \cite{LAMA_Paper}, the classical MMSE-SIC \cite{MMSE_SIC2}, the EP approximation (EPA) \cite{EP2019}, the  EP-NSA \cite{EP-NSA2018}, and the near optimal EP detectors \cite{Jespedes-TCOM14} in a  Rician fading channel. {The}  Rician fading model generalizes the Rayleigh fading model by considering the line-of-sight (LOS) between the transmitter and receiver. A Rician channel from the $k$-th transmit antenna to   the $n$-th receive antenna is characterized by $h_{n,k} \sim \mathcal{N}\left(\mu_{\rm Ric}, \sigma_{\rm Ric}^2  \right)$, where $\mu_{\rm Ric} = \sqrt{\frac{\phi}{2(\phi+1)}}$, $\sigma_{\rm Ric}^2= \frac{1}{2(\phi+1)}$, and $\phi$ represents {the}  Rician factor, i.e. the ratio of power of the LOS component to the power of the scattered components \cite{WirelessCommBook}.  Note that when $\phi=0$, a Rician fading channel reduces to a Rayleigh fading channel. The BER
performance of our proposed detectors in a Rician fading channel is depicted in Fig. \ref{F9b}. The proposed B-PIC-DSC detector exhibits a similar performance to the AMP detector for  the Rician factor $\phi>1.5$. The proposed IB-PIC-DSC detector outperforms the EPA, EP-NSA, and MMSE-SIC detectors for $\phi<3$. As shown in Section \ref{sComplAnalysis}, the IB-PIC-DSC detector has  a similar computational complexity as the EPA detector and a much lower complexity than the EP-NSA and MMSE-SIC detectors.

\subsection{Polar-Coded M-MIMO Receiver}
\label{sNumericPolar}

The error-correction performance and the complexity of the proposed polar-coded M-MIMO receiver are investigated for the case of Rayleigh fading channel with 4-QAM modulation. 5G New Radio polar codes \cite{polarNR2018} of the length $\eta=256,512$ and the code rate $R=1/4,1/2$ are considered. The number $\kappa$ of information bits of a polar code can be computed as $\kappa=R\cdot\eta+s$, where $s$ is the CRC length. In case of $\eta=256,512$ and $R=1/4,1/2$, polar codes with CRC of length $11$ (uplink scenario) demonstrate a much better performance than polar codes with CRC of length $24$ (downlink scenario), therefore we provide results only for the uplink scenario. 

\begin{figure*}
\centering
\subfloat[$ \eta=256, R=\frac{1}{2}, N=64,  K=16$ ]
{\includegraphics[scale=0.34]{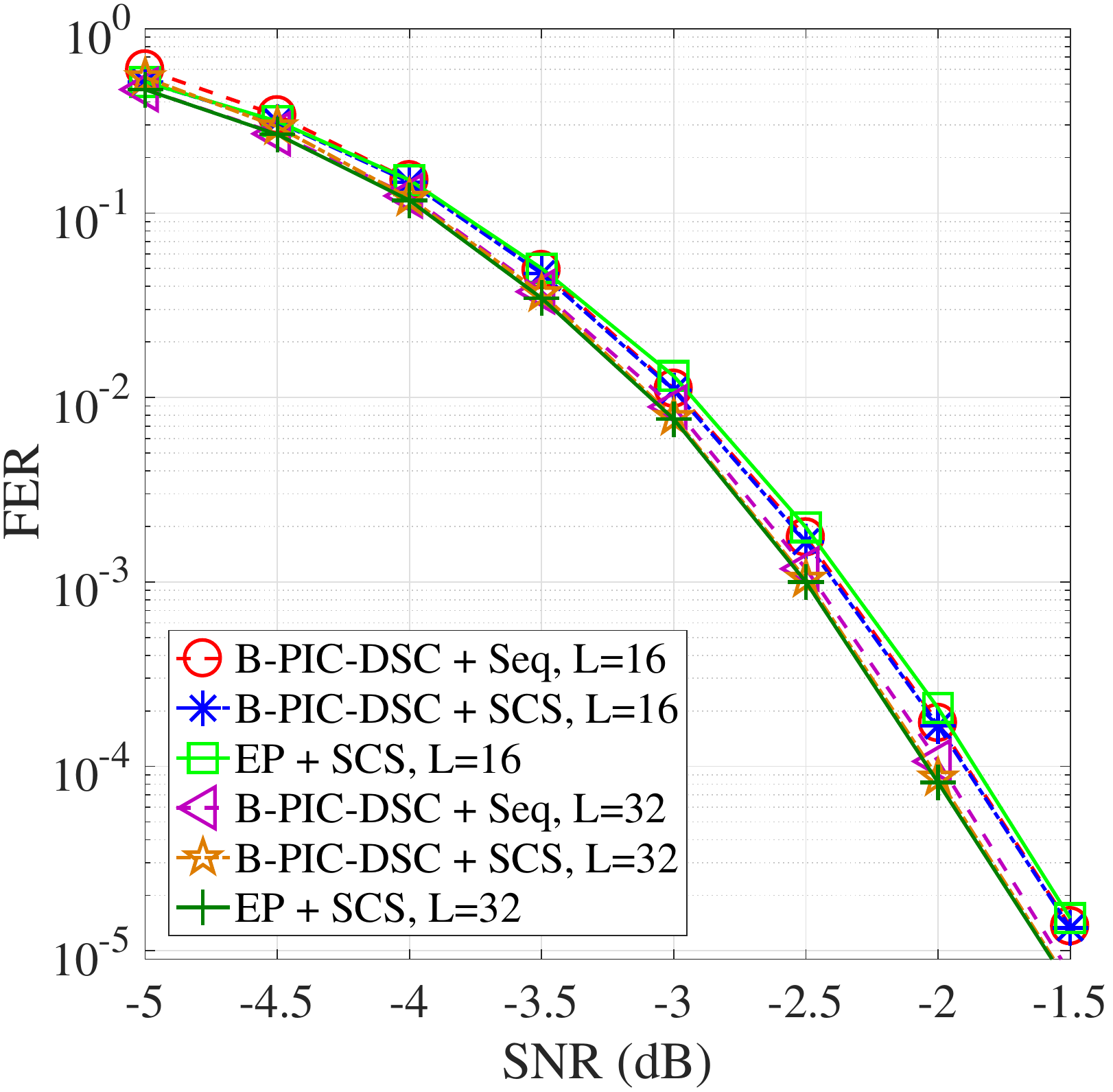}}\hfill
\centering
\subfloat[$ \eta=256, R=\frac{1}{4}, N=K=128$ ]
{\includegraphics[scale=0.34]{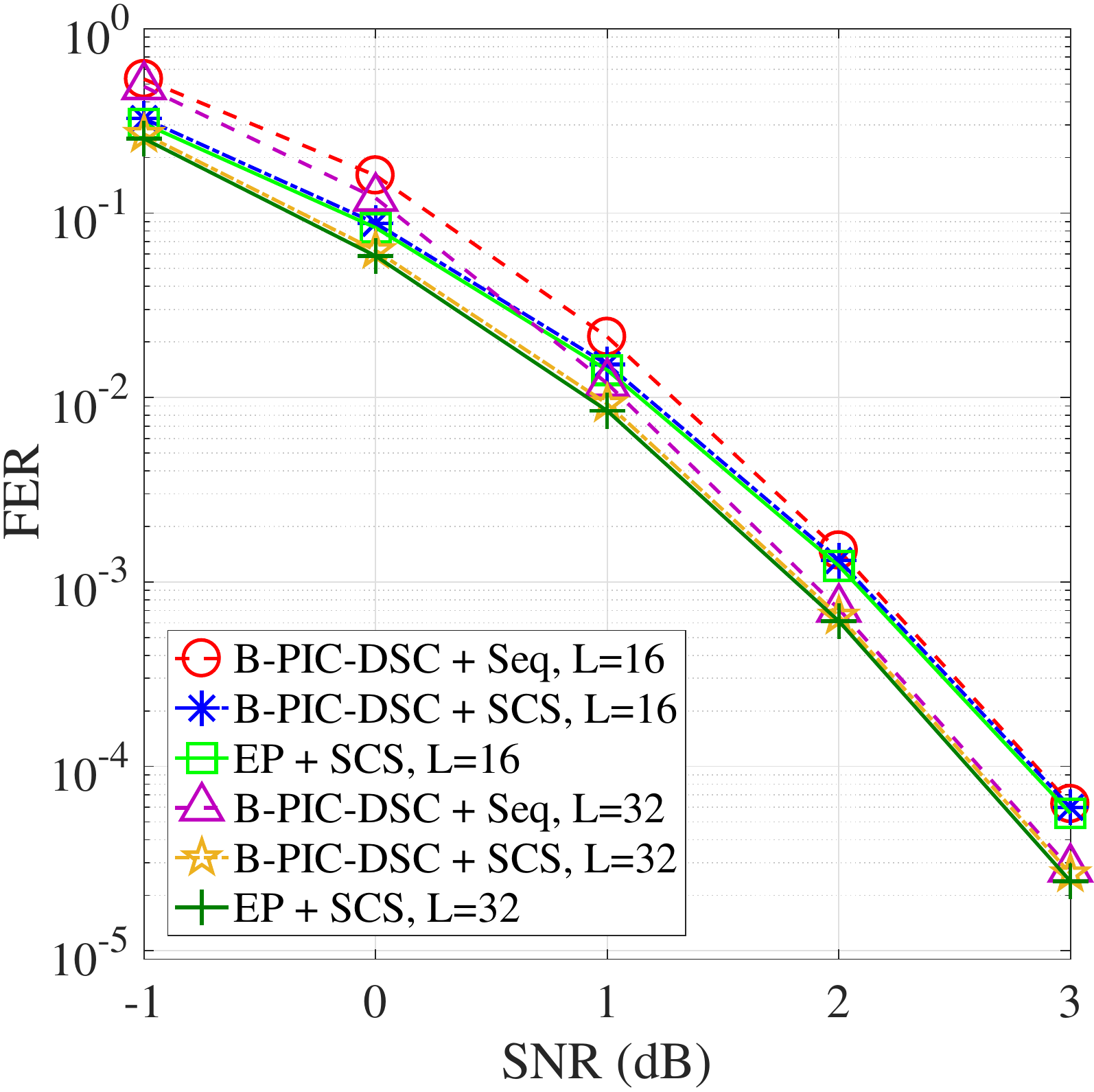}}\hfill
\centering
\subfloat[$ \eta=512$, $R=\frac{1}{2}, N=64$, $K=16$ ]
{\includegraphics[scale=0.34]{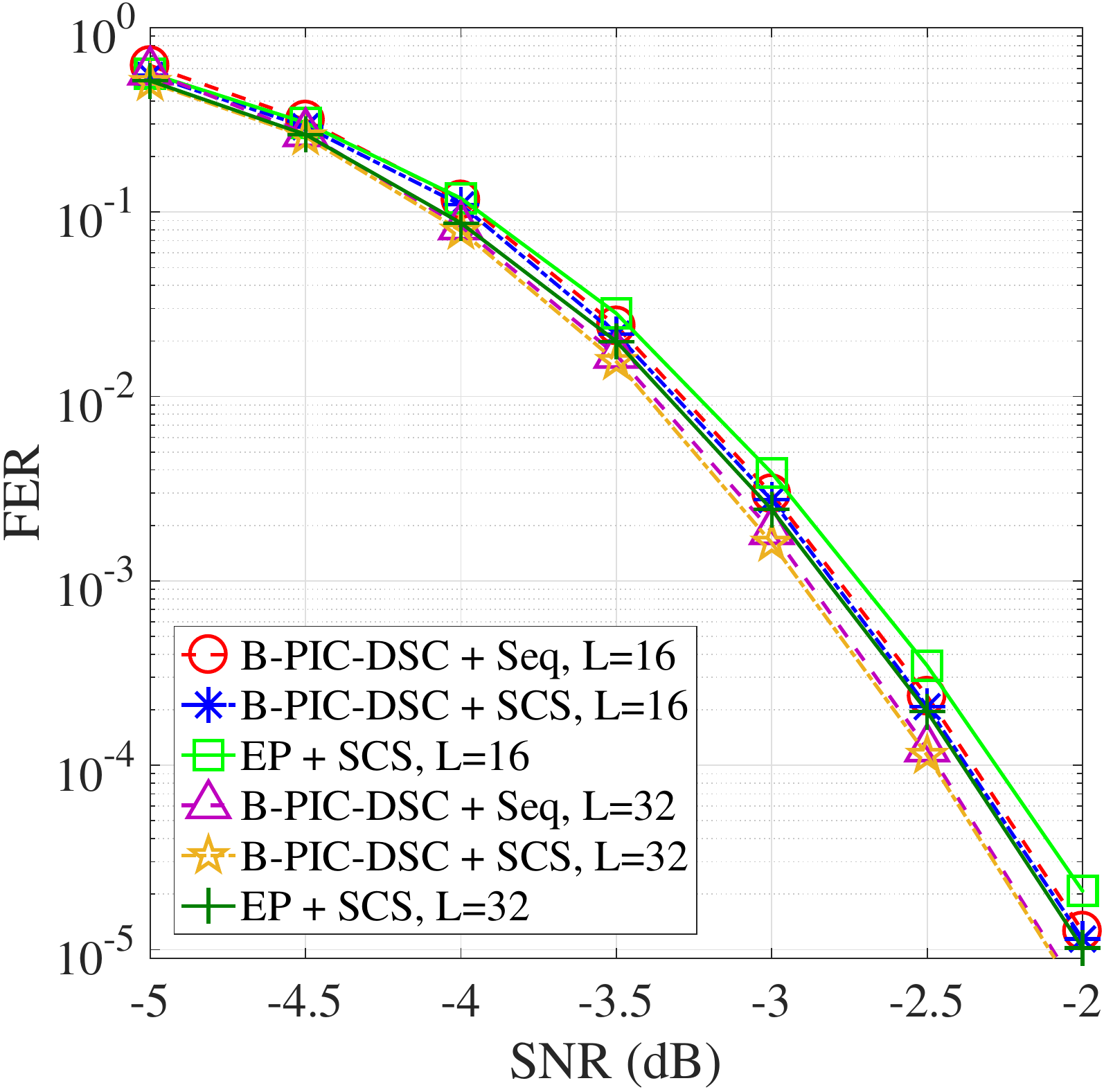}}
\caption{The FER performance comparison of the polar-coded M-MIMO receivers}
\label{FER_results}
\end{figure*}

A comparison is provided for the following polar-coded M-MIMO receivers: 
\begin{itemize}
\item The proposed receiver (see Section \ref{sPolar}), which is based on the proposed B-PIC-DSC detector (see Sections \ref{sDetector},\ref{sImprDetector}), the variance analysis in Appendix \ref{Ap.1} and the sequential decoder \cite{miloslavskaya2014sequential,Trifonov2018ASF}, labeled as ``B-PIC-DSC+Seq'';
\item A receiver comprising the proposed B-PIC-DSC detector and  the  SCS decoder \cite{niu2012crcaided}, labeled as ``B-PIC-DSC+SCS'';
\item A receiver comprising the EP  detector  \cite{Jespedes-TCOM14}  (near ML detector) and  the  SCS decoder \cite{niu2012crcaided}, labeled as ``EP+SCS'';
\end{itemize}  
where the decoding is performed with the list size $L=16,32$. Note that  the  SCS demonstrate the same performance as widely known  SCL \cite{tal2011list} with the same list size.

\begin{figure*}
\centering
\subfloat[$ \eta=256, R=\frac{1}{2}, N=64, K=16$]
{\includegraphics[scale=0.33]{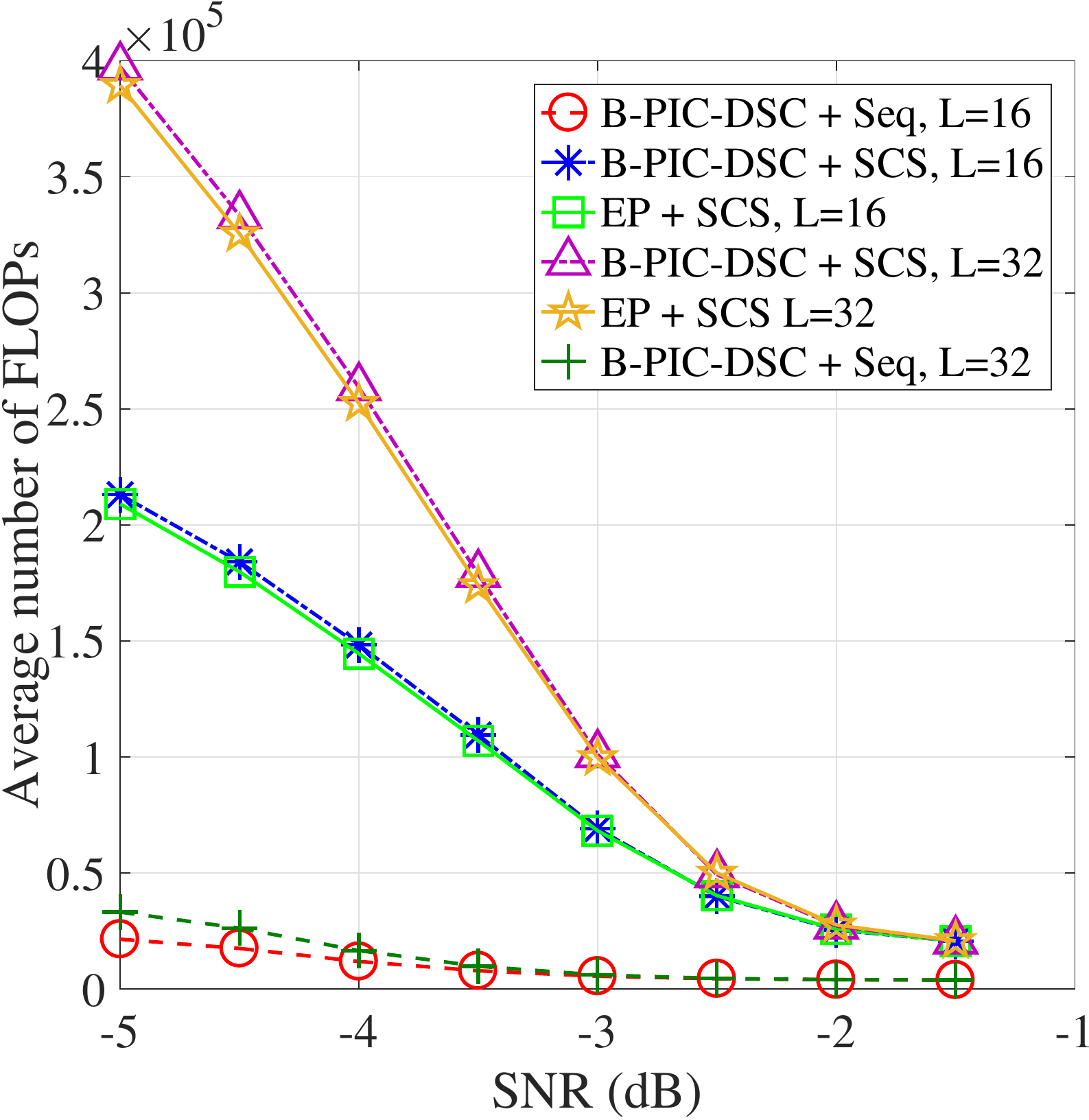}}\hfill
\centering
\subfloat[$ \eta=256, R=\frac{1}{4}, N=K=128$]
{\includegraphics[scale=0.33]{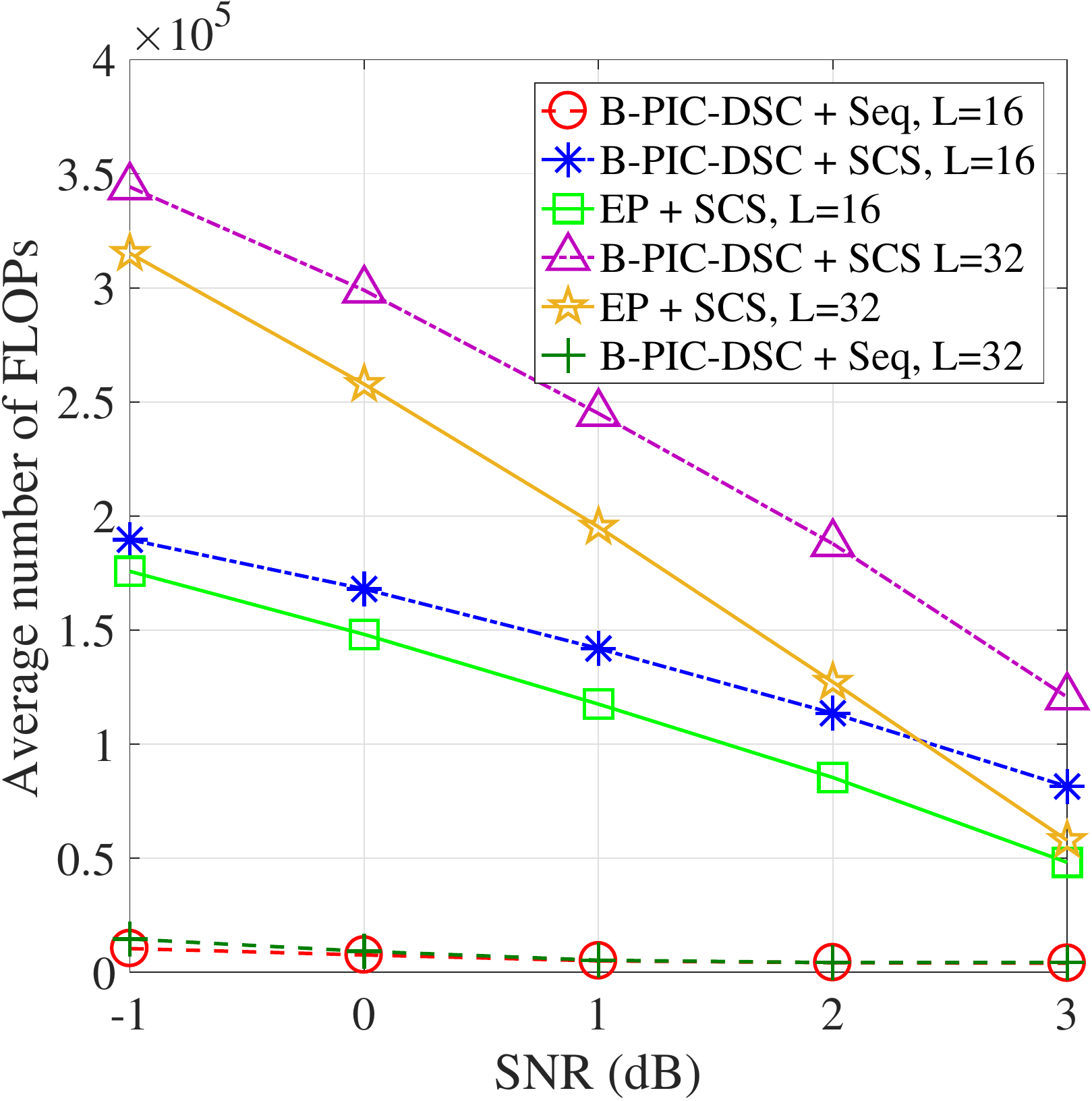}}\hfill
\centering
\subfloat[$ \eta=512$, $R=\frac{1}{2}, N=64$, $K=16$]
{\includegraphics[scale=0.33]{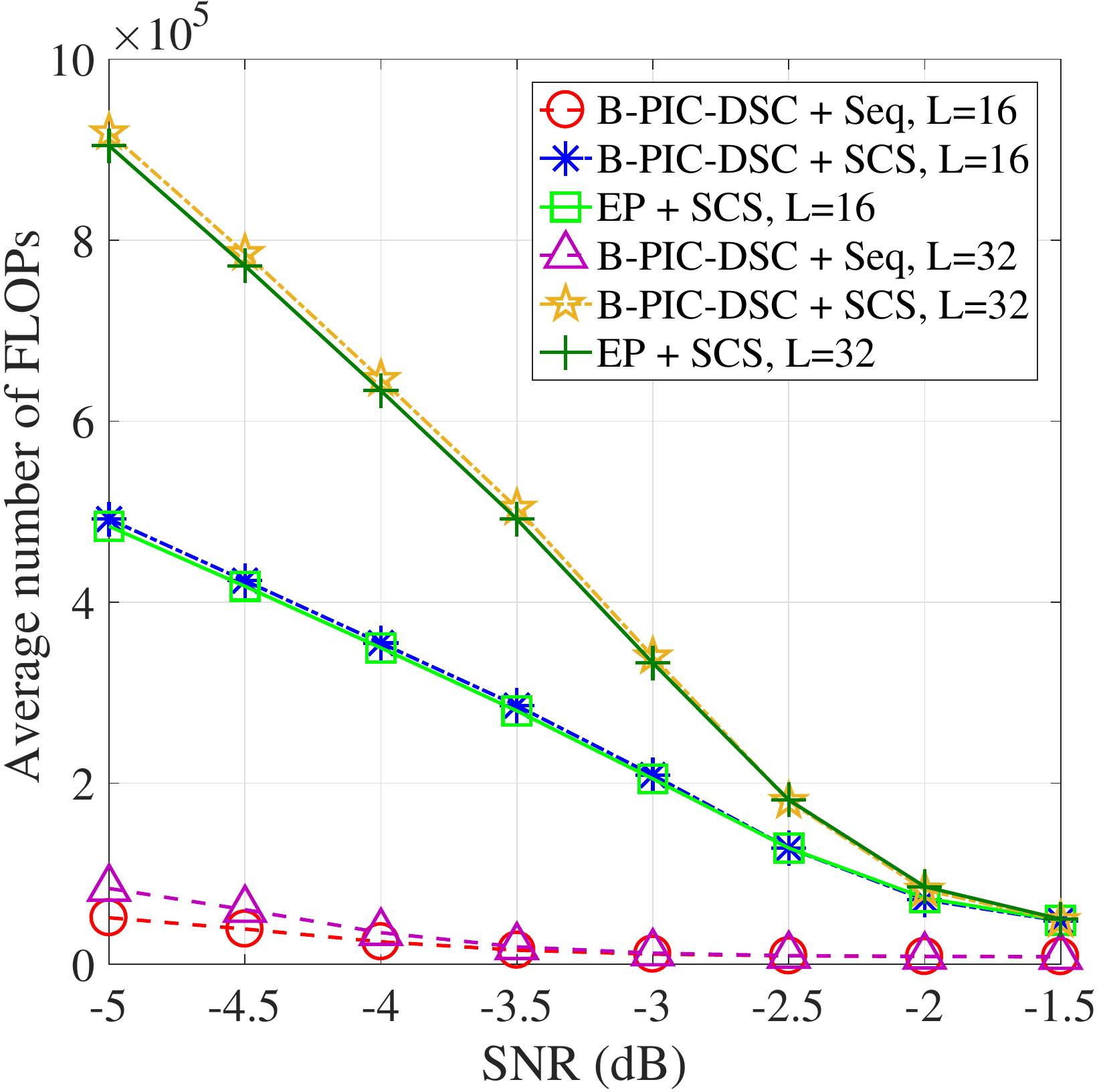}}
\caption{The average number of FLOPs comparison of the polar-coded M-MIMO receivers}
\label{Iter_results}
\end{figure*}

\begin{figure*}
\centering
\subfloat[ $ \eta=256, R=\frac{1}{2}, N=64, K=16$]
{\includegraphics[scale=0.33]{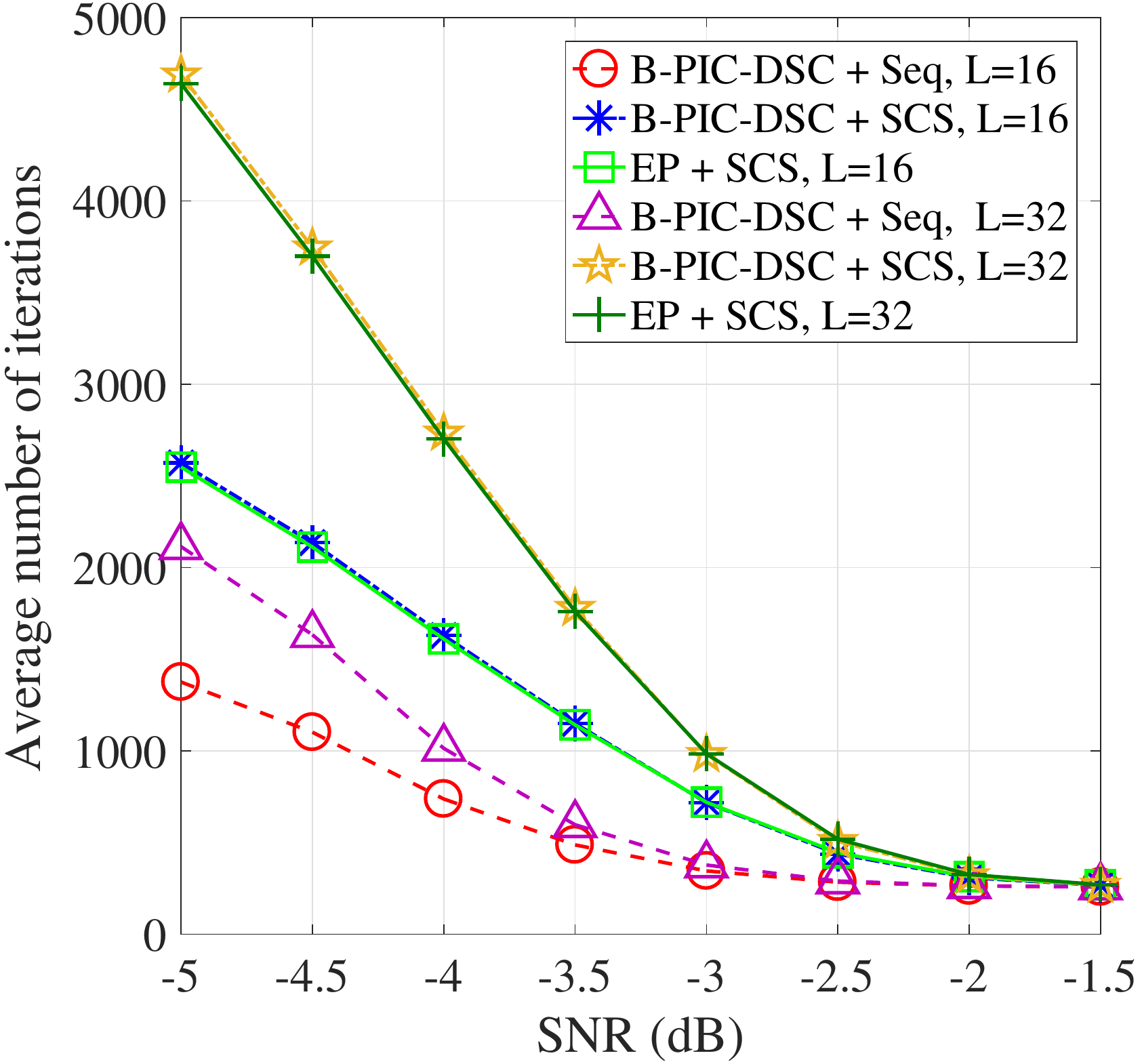}}\hfill
\centering
\subfloat[$ \eta=256, R=\frac{1}{4}, N=K=128$ ]
{\includegraphics[scale=0.33]{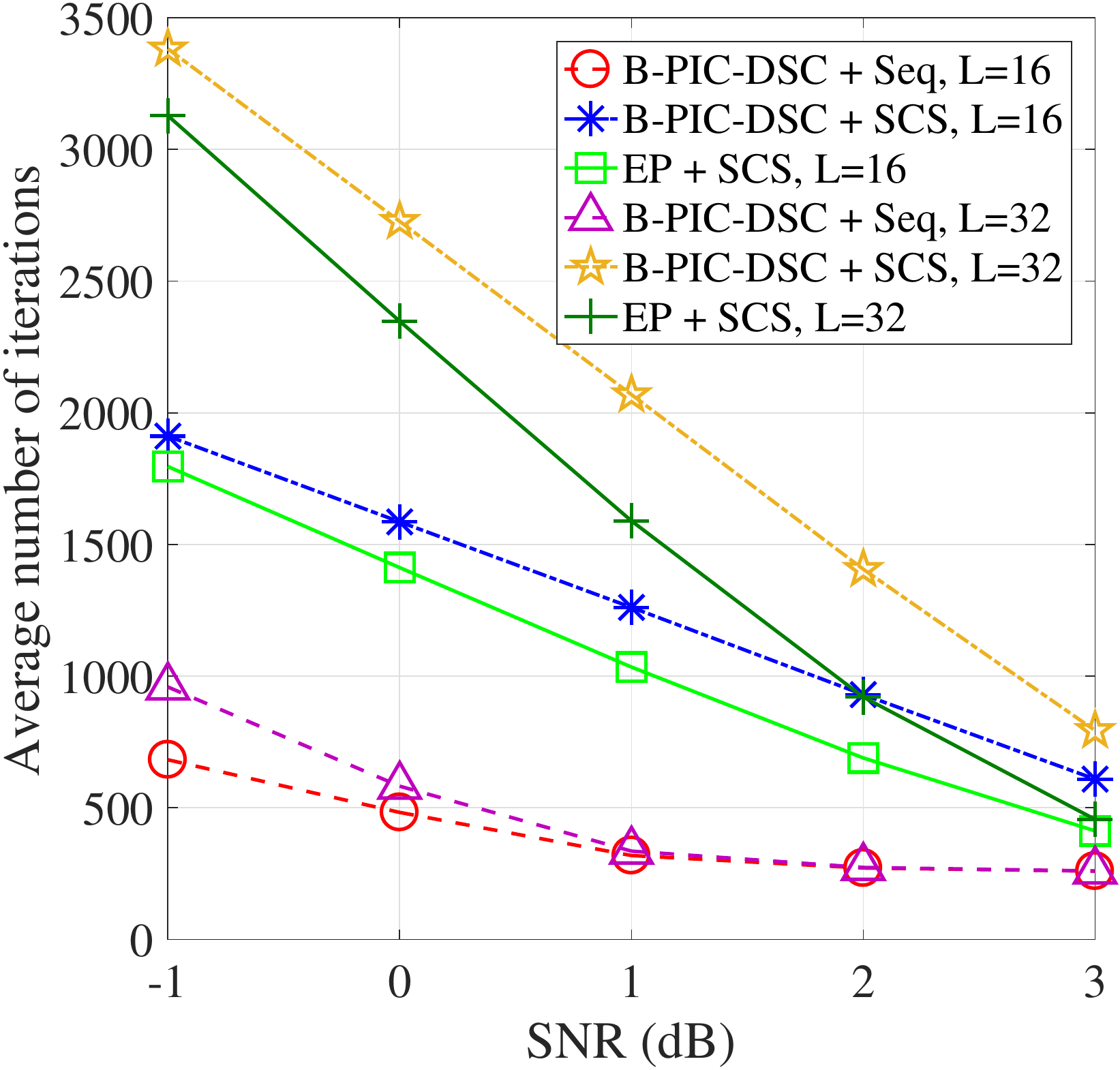}}\hfill
\centering
\subfloat[ $ \eta=512, R=\frac{1}{2}, N=64, K=16$]
{\includegraphics[scale=0.33]{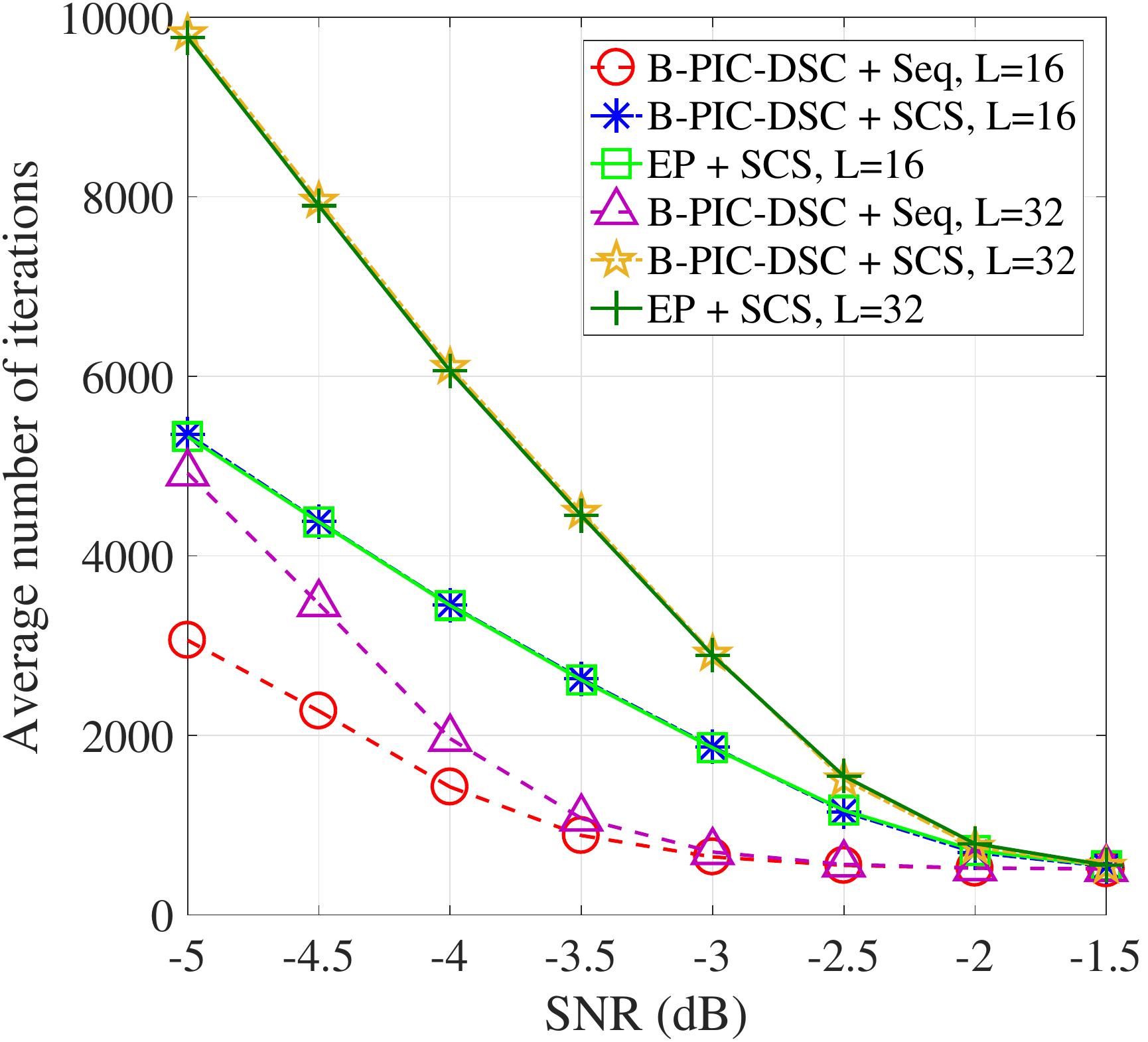}}
\caption{The average number of iterations comparison of the polar-coded M-MIMO receivers}
\label{FLOPs_results}
\end{figure*}

The performance comparison in terms of the FER is shown in Fig. \ref{FER_results}. It can be seen that ``B-PIC-DSC+Seq'', ``B-PIC-DSC+SCS'' and ``EP+SCS'' demonstrate a similar performance. Note that the performance improves with the list size $L$ at the expense of an increase in the  complexity. The  complexity is characterised by the number of iterations and the number of floating point operations (FLOPs). Fig. \ref{Iter_results} presents the average number of iterations per decoding attempt. The proposed ``B-PIC-DSC+Seq'' requires significantly fewer iterations than ``B-PIC-DSC+SCS'' and ``EP+SCS'' due to the presence of the bias function in the sequential decoder (see \ref{sPolarSeq}), which is absent in  the SCS. Fig. \ref{FLOPs_results} shows the average number of FLOPs per decoding attempt. In case of  the  SCS, these FLOPs are additions, comparisons and multiplications. The sequential decoder does not require floating point multiplications due to simplified recursive computations in log-domain. There are two reasons for the significant reduction of the FLOP number demonstrated by the sequential decoder: reduced number of iterations shown in Fig. \ref{Iter_results} and reduced complexity of each iteration due to elimination of the multiplication operations. Thus, the proposed M-MIMO receiver ``B-PIC-DSC+Seq'' significantly outperforms other receivers.

\section{Conclusion}

We propose  novel Bayesian based PIC detectors and use them to design a polar-coded M-MIMO receiver to achieve a high reliability with a low processing delay. Both detectors are near optimal in the sense that the achieved BERs are very close to that of the optimal ML and EP detectors. On the other hand, the complexity of the  B-PIC-DSC detector increases linearly with the number of receive antennas while the complexity of the IB-PIC-DSC is much lower than the EP detector. The convergence rate of the IB-PIC-DSC detector is better than that of the B-PIC-DSC detector implying that the low complexity signal processing of the B-PIC-DSC detector comes at the expense of a slightly increased number of iterations.
 Simulation results show that,  the proposed detectors in Rayleigh fading channel can provide approximately a $1$ dB  performance gain compared to the AMP detector, in uncoded systems with one order of magnitude lower complexity compared to  the  EP detector. The proposed B-PIC-DSC detector has been integrated with the sequential decoder for polar codes to realize a low complexity polar-coded M-MIMO receiver. This integration was enabled by the closed-form variance approximation of the log-likelihood ratios calculated  by  the B-PIC-DSC  detector. Simulation results for polar codes from the 5G New Radio standard show that the proposed M-MIMO receiver ensures one order magnitude lower  complexity and a similar FER performance compared to other receivers. One of possible directions for future work is integration of the proposed detector with a neural network (NN) technique to further improve the BER performance.

\begin{appendices}
\section{Derivation of \eqref{eA1_a01}}
\label{Ap.Var_PIC}
The $k$-th symbol estimate from PIC scheme $x_{{\rm PIC},k}^{(t)}$ can be expanded by substituting \eqref{eII_1} to \eqref{eA1_a02}, 
\begin{multline}\label{Ap_PIC_ori}
x_{{\rm PIC},k}^{(t)} = x_k + \frac{ \sum_{j=1,j \neq k}^{K} \sum_{n=1}^{N}h_{n,k}^* h_{n,j} \left( x_j - x_{{\rm PIC},j}^{(t-1)}\right)}{\sum_{n=1}^{N}h_{n,k}^* h_{n,k}} \\
 +\frac{\sum_{n=1}^{N}h_{n,k}^*\varepsilon_n}{\sum_{n=1}^{N}h_{n,k}^* h_{n,k}}
\end{multline}
From \eqref{Ap_PIC_ori}, we can obtain the variance $\Sigma^{(t)}_k \triangleq \mathrm{Var} \left[(x^{(t)}_{{\rm PIC},k}-x_{k})\right] $ as follows,
\begin{multline}
\Sigma_k^{(t)} =\mathrm{Var} \left[  \frac{ \sum_{j=1,j \neq k}^{K} \sum_{n=1}^{N}h_{n,k}^* h_{n,j} \left( x_j - x_{{\rm PIC},j}^{(t-1)}\right)}{\sum_{n=1}^{N}h_{n,k}^* h_{n,k}} \right]  \\
 + \mathrm{Var} \left[ \frac{\sum_{n=1}^{N}h_{n,k}^*\varepsilon_n}{\sum_{n=1}^{N}h_{n,k}^* h_{n,k}} \right].
 \end{multline}
 According to the definition of variance,
\begin{align} \label{Ap_PICvar1}
\Sigma_k^{(t)} = & \Ex \left[ \left(  \frac{\sum_{j=1,j \neq k}^{K} s_j \left( x_j - x_{{\rm PIC},j}^{(t-1)}\right)}{\sum_{n=1}^{N}h_{n,k}^* h_{n,k}}   \right)^2 \right] \notag  - \\ 
& \left( \Ex \left[  \frac{\sum_{j=1,j \neq k}^{K} s_j \left( x_j - x_{{\rm PIC},j}^{(t-1)}\right) }{\sum_{n=1}^{N}h_{n,k}^* h_{n,k}}  \right]\right)^2 \notag \\
 &+ \Ex   \left[ \left( \frac{\sum_{n=1}^{N}h_{n,k}^*\varepsilon_n}{\sum_{n=1}^{N}h_{n,k}^* h_{n,k}} \right)^2 \right] - \left(\Ex  \left[ \frac{\sum_{n=1}^{N}h_{n,k}^*\varepsilon_n}{\sum_{n=1}^{N}h_{n,k}^* h_{n,k}} \right]\right)^2,
\end{align}
where  $s_j =\sum_{n=1}^{N} h_{n,k}^*h_{n,j}$. As we take the expectation over the noise $\varepsilon_n$, the last term in \eqref{Ap_PICvar1} is zero and $\Ex  \left[  \left( \sum_{n=1}^{N}h_{n,k}^*\varepsilon_n \right)^2\right] = \Ex  \left[ \sum_{n=1}^{N} ( h_{n,k}^*\varepsilon_n )^2\right]$. 
Substituting $\Ex [(\varepsilon_n)^2] = \sigma^2$, $\Sigma^{(t)}_k$ can be expressed as follows,
\begin{multline} \label{Ap_PICvar2}
\Sigma_k^{(t)} = \frac{1}{\left( \sum_{n=1}^{N}h_{n,k}^* h_{n,k}\right)^2} 
\Bigg\{ \sum_{j=1,j \neq k}^{K} s_j^2 \Bigg( \Ex \Big[  \big(  x_j - x_{{\rm PIC},j}^{(t-1)} \big)^2  \Big] \\ 
- \Big( \Ex \big[x_j - x_{{\rm PIC},j}^{(t-1)}\big] \Big)^2 \Bigg)  +  \sum_{n=1}^N  \big( h_{n,k}^*h_{n,k} \big) \sigma^2  \Bigg\}.
\end{multline}

To simplify the derivation, we assume the DSC coefficients are equal to unity meaning that we do not perform any linear combination of the current and previous symbol estimates and thus $x_{{\rm PIC},j}^{(t-1)} = \hat{x}_j^{(t-1)}$, where $\hat{x}_j^{(t-1)} $ is  given in \eqref{eA1_b01}. 
From \eqref{eA1_b02}, we can obtain $ V_j^{(t-1)} =  \Ex \left[ \left(  x_j - \hat{x}_j^{(t-1)}  \right)^2  \right] - \left( \Ex \left[x_j - \hat{x}_j^{(t-1)}\right] \right)^2 $. Therefore,  \eqref{Ap_PICvar2} can be rewritten as
\begin{align}
\Sigma_k^{(t)} &= \frac{1}{\left( \sum_{n=1}^{N}h_{n,k}^* h_{n,k}\right)^2} \left( \sum_{\substack{j=1 \\  j\neq k}}^{K} s_j^2 V_j^{(t-1)}  +  \sum_{n=1}^N  \left( h_{n,k}^*h_{n,k} \right) \sigma^2 \right)  \\ \notag
&\approx \frac{\sigma^2}{\sum_{n=1}^{N}h_{n,k}^* h_{n,k}},
\end{align}
where the approximation is given in \cite{LBL_WCNC}. 
This complete the proof.

\section{Derivation of \eqref{SINR_BSO}}
\label{Ap.1}

First, we substitute $\qy=\qH\qx+\qvarepsilon$ into  \eqref{eA1_a02} i.e.
\begin{equation}\label{A_1}
x_{{\rm PIC},k}^{(t)}= x_k+ \frac{\sum_{j=1,j\neq k}^{K}\qh_{k}^H \qh_{j}\left( x_j - x_{{\rm PIC},j}^{(t-1)} \right) 
+ \qh_k^H \qvarepsilon}{\| \qh_k \|^2}.
\end{equation}
 From \eqref{A_1}, the ergodic achievable sum rate, $\mathbf{C}$ can then be derived as \cite{Caire2015}
\begin{flalign}
\mathbf{C} = \Ex \left[   \sum_{k=1}^{K} {\rm log}_2  \left\lbrace  1+ \frac{\|\qh_k\|^2}{\sum_{j=1,j \neq k}^{K} \frac{  |\qh_k^T\qh_j|^2}{\|\qh_k\|^2} V_j^{(t-1)} +  \sigma^2}   \right\rbrace \right], \label{A_2}
\end{flalign} 
where $V_j^{(t-1)} = \Ex \left[ |x_j - x_{{\rm PIC},j}^{(t-1)}|^2 \right]$ and the expectation in \eqref{A_2} is taken with respect to channel gain matrix $\qH=[\qh_1,\dots,\qh_K]$. 
Note that $x_{{\rm PIC},j}^{(t-1)} $ is identical to $\hat{x}^{(t-1)}_j $ as we set the coefficients of the linear combination, $\qrho_{\rm DSC}^{(t-1)}$ in \eqref{DSC_coef} equal to unity. 
Lemma 4 in \cite{Caire2015} shows that when the number of receive antennas grow very large, we can approximate the last term in \eqref{A_2} as 
\begin{flalign}
\mathbf{C}  & \approx  \sum_{k=1}^{K} {\rm log}_2  \left\lbrace  1+ \frac{\Ex \left[ \|\qh_k\|^2 \right]}{\sum_{j=1,j \neq k}^{K}  \frac{  \Ex \left[ |\qh_k^T\qh_j|^2\right]}{\|\qh_k\|^2} V_j^{(t-1)} +  \sigma^2  }   \right\rbrace \label{SINR}    \\ \notag
&= \sum_{k=1}^{K} {\rm log}_2    \Bigg\{  1+ \underbrace{\frac{1}{\frac{(K-1) V_j^{(t-1)}+  \sigma^2}{N}}}_{\text{SINR}}    \Bigg\} 
\end{flalign}
The last term in \eqref{SINR} refers to  the SINR of the BSO module, denoted by $\frac{1}{v^{(t)}}$. Therefore,  $v^{(t)}$ in \eqref{SINR_BSO} can be re-expressed as follows,
 \begin{equation}\label{A_4}
v^{(t)} =  \frac{(K-1)}{N}V_j^{(t-1)} +  \frac{\sigma^2}{N}.
\end{equation}
At the first iteration, we set $V_j^{(0)}=1$, and thus $v^{(0)} $ is identical to that of the MSE expression of the matched filter detector given in \cite{J.Hoydis2013}. This completes the derivation.

\end{appendices}

\section*{Acknowledgment}

This research was supported by the research training program stipend  from The University of Sydney. The work of Branka Vucetic was supported in part by the Australian Research Council Laureate Fellowship grant number FL160100032.

\ifCLASSOPTIONcaptionsoff
  \newpage
\fi

%
%

\bibliographystyle{IEEEtran} 
\bibliography{IEEEabrv,myBib}

\begin{IEEEbiography}
 [{\includegraphics[width=1in,height=1.25in,clip,keepaspectratio]{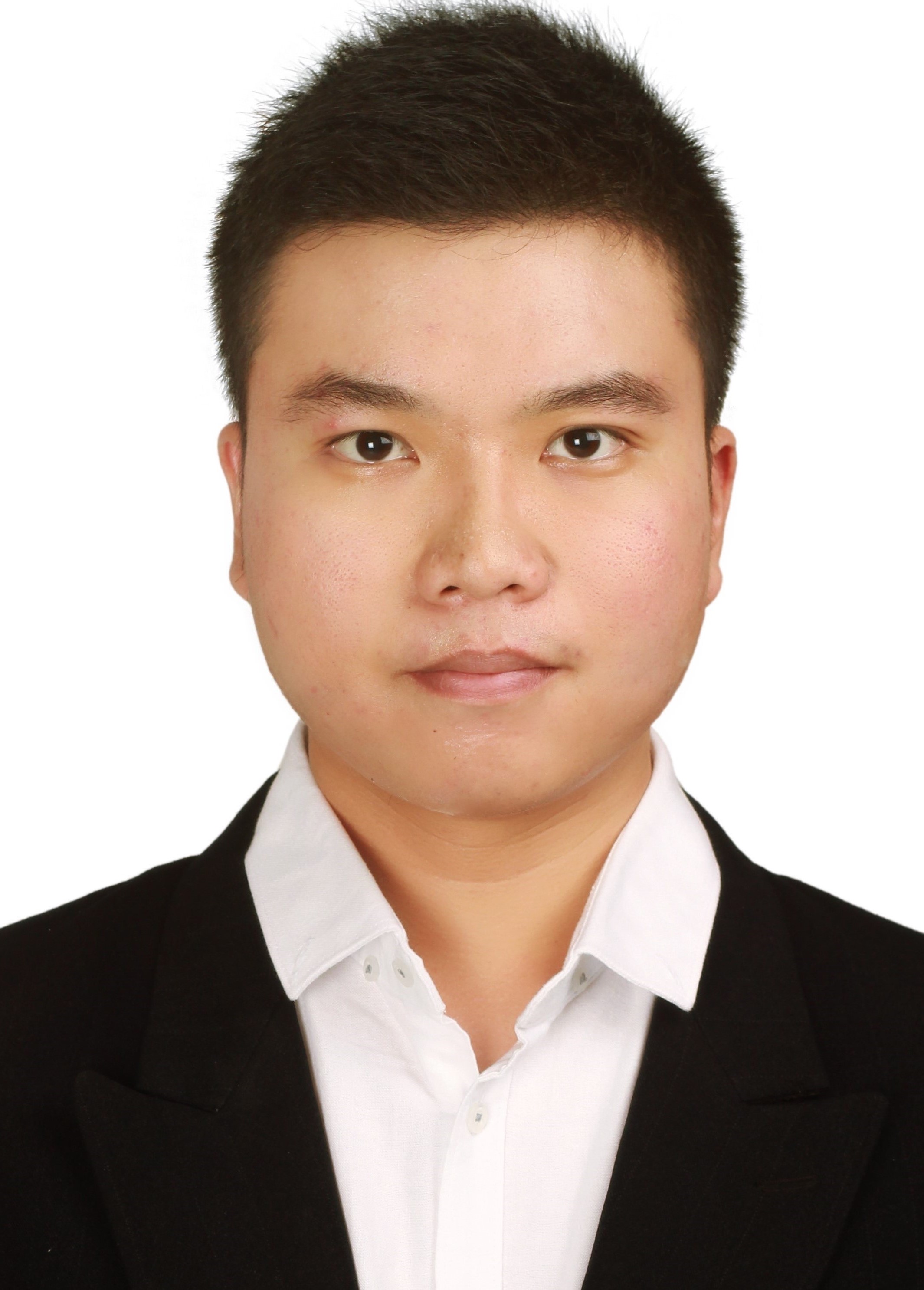}}]{Alva Kosasih}(S’19)
Alva Kosasih received the B.Eng. and M.Eng. degrees both in electrical engineering from Brawijaya University, Indonesia, in 2013 and 2017,
respectively; and the M.S. degree in communication engineering from National Sun Yat-sen University, Taiwan, in 2017. He is currently pursuing the Ph.D degree in the School of Electrical and Information Engineering, the University of Sydney, Australia. His research interests
include massive MIMO systems and the application of Bayesian and machine learning in wireless communications
\end{IEEEbiography}

\begin{IEEEbiography}
 [{\includegraphics[width=1in,height=1.25in,clip,keepaspectratio]{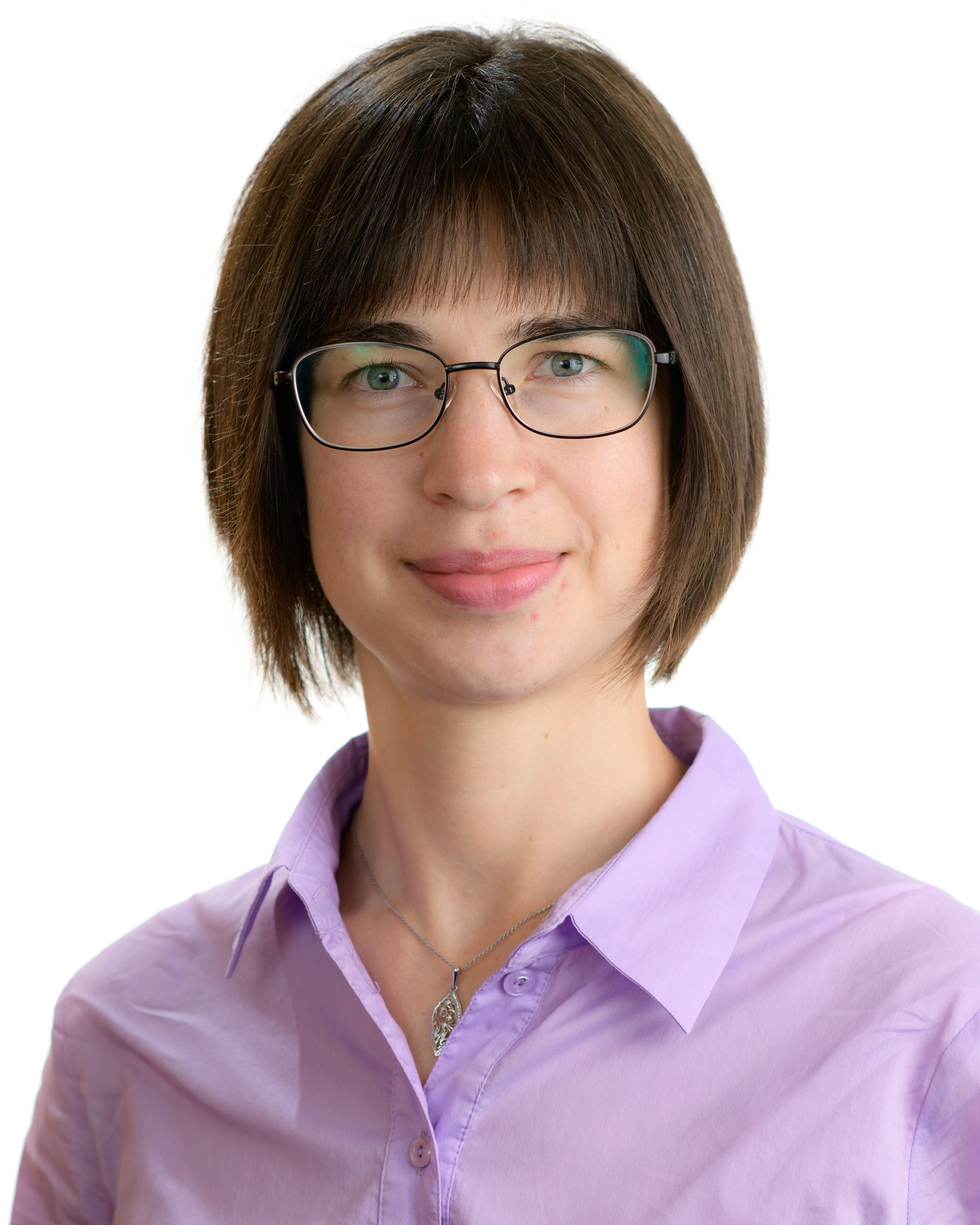}}]{Vera Miloslavskaya}
received B.Sc., M.Sc. and PhD degrees from Peter the Great St. Petersburg Polytechnic University (SPbPU) in 2010, 2012 and 2015, respectively. Her
research interests include coding theory and its applications in telecommunications and storage systems. She is currently a Postdoctoral Research Associate in Telecommunications in the School of Electrical and Information Engineering at the University of Sydney.
\end{IEEEbiography}

\begin{IEEEbiography}
[{\includegraphics[width=1in,height=1.25in,clip,keepaspectratio]{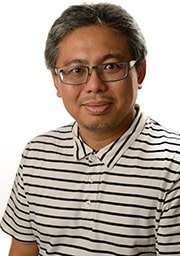}}]{Wibowo Hardjawana }(M'09) 
received the Ph.D. degree in electrical engineering from The University of Sydney, Australia, in 2009. He was an Australian Research Council Discovery Early Career Research Award Fellow and is now Senior Lecturer with the School of Electrical and Information Engineering, The University of Sydney. Prior to that he was Assistant Manager at Singapore Telecom Ltd, managing core and radio access networks. His current research interests are in 5/6G cellular radio access and wireless local area networks, with focuses in system architectures, resource scheduling, interference, signal processing and the development of corresponding standard-compliant prototypes.

\end{IEEEbiography}

\begin{IEEEbiography}
 [{\includegraphics[width=1in,height=1.25in,clip,keepaspectratio]{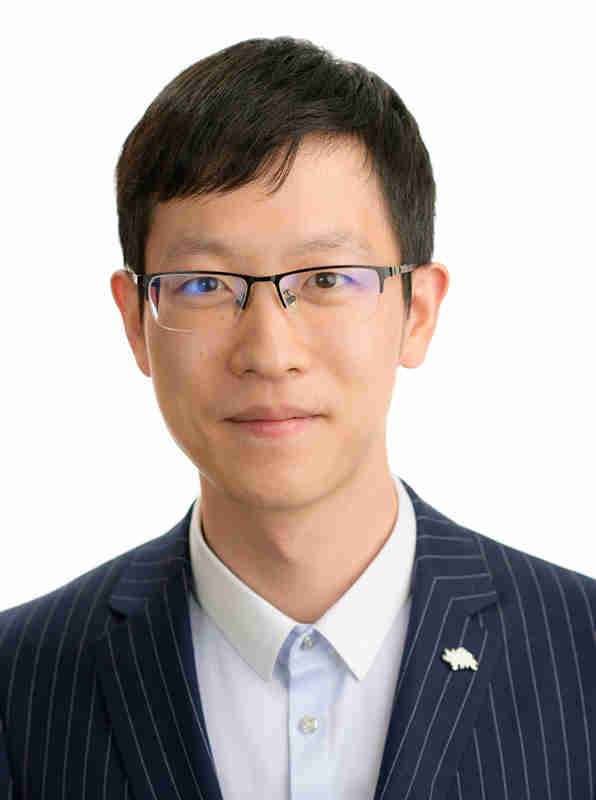}}]{Changyang She}(S'12-M'17) 
received his B. Eng degree in Honors College (formerly School of Advanced Engineering) of Beihang University (BUAA), Beijing, China in 2012 and Ph.D. degree in School of Electronics and Information Engineering of BUAA in 2017. From 2017 to 2018, he was a postdoctoral research fellow at Singapore University of Technology and Design. From 2018 to 2021, he was a postdoctoral research associate at the University of Sydney. Now, he serves as the Australian Research Council Discovery Early Career Research Award (ARC DECRA) Fellow at the University of Sydney. His research interests lie in the areas of ultra-reliable and low-latency communications, deep learning in wireless networks, mobile edge computing, and Internet-of-Things. 
\end{IEEEbiography}

\begin{IEEEbiography}
 [{\includegraphics[width=1in,height=1.25in,clip,keepaspectratio]{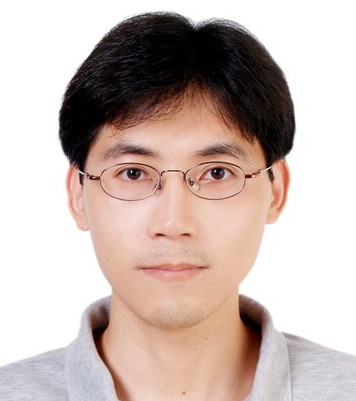}}]{Chao-Kai Wen}(S'00--M'04-SM'20)
received a Ph.D. degree from the Institute of Communications Engineering, National Tsing Hua University, Taiwan, in 2004.
He was with Industrial Technology Research Institute, Hsinchu, Taiwan, and MediaTek Inc., Hsinchu, Taiwan, from 2004 to 2009, where he was engaged in broadband digital transceiver design. Since 2009, he joined the Institute of Communications Engineering, National Sun Yat-sen University, Kaohsiung, Taiwan, where he is currently a Professor. His research interests center around optimization in wireless multimedia networks.
\end{IEEEbiography}

\begin{IEEEbiography}
 [{\includegraphics[width=1in,height=1.25in,clip,keepaspectratio]{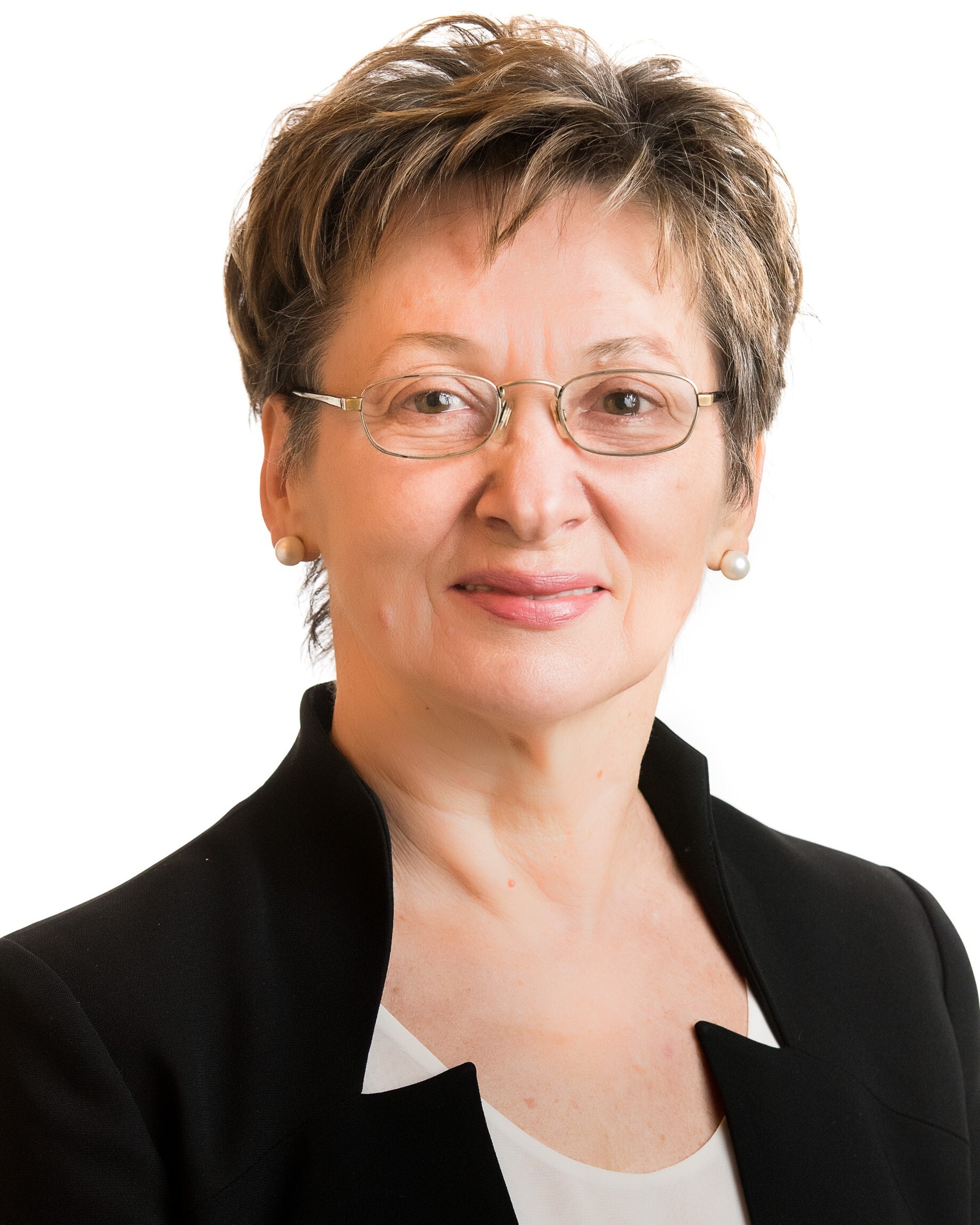}}]{Branka Vucetic}(Life Fellow, IEEE)
is an ARC Laureate Fellow and Director of the Centre of Excellence for IoT and Telecommunications at the University of Sydney. 
Her current research work is in wireless networks and the Internet of Things. In the area of wireless networks, she works on ultra-reliable low-latency communications (URLLC) and system design for millimetre wave frequency bands. In the area of the Internet of Things, Vucetic works on providing wireless connectivity for mission critical applications. Branka Vucetic is a Fellow of IEEE, the Australian Academy of Technological Sciences and Engineering and the Australian Academy of Science. The work of Branka Vucetic was supported in part by the Australian Research Council Laureate Fellowship grant number FL160100032.
\end{IEEEbiography}





\end{document}